\documentclass[preprint,showpacs,showkeys]{revtex4}
\usepackage{graphicx}
\usepackage{subfigure}
\usepackage{amsmath,amssymb}
\usepackage{psfrag}

\usepackage[latin1]{inputenc}
\usepackage[T1]{fontenc}

\newcommand{\de}{\mbox{d}}
\newcommand{\eref}[1]{(\ref{#1})}
\newcommand{\ii}{\text{i}}

\begin{document}

\title{Traveling waves and Compactons in Phase Oscillator Lattices}
\date{\today}

\author{Karsten Ahnert}
\author{Arkady Pikovsky}
\affiliation{Department of Physics, Potsdam University, Potsdam, Germany}

\begin{abstract}
We study waves in a chain of dispersively coupled phase oscillators.  Two
approaches -- a quasi-continuous approximation and an iterative numerical
solution of the lattice equation -- allow us to characterize different types
of traveling waves: compactons, kovatons, solitary waves with exponential
tails as well as a novel type of semi-compact waves that are compact from one
side. Stability of these waves is studied using numerical simulations of the
initial value problem.
\end{abstract}


\maketitle

\section{\label{sec:intro}Introduction}

{\bf The topic of this paper unifies two principal directions of nonlinear
  science: coupled self-sustained oscillators and soliton theory. Coupled
  autonomous self-sustained oscillators appear in different fields of science,
  they demonstrate a variety of fascinating phenomena. In this study we
  demonstrate that particularities of the coupling for a rather simple setup
  -- all oscillators are identical, periodic, and form a regular lattice with
  a nearest neighbor coupling -- can lead to highly nontrivial wave structures.
  Remarkably, a dispersive coupling of the phases of dissipative oscillators
  results in a simple lattice equation which is equivalent to a Hamiltonian
  one. We show that different types of traveling waves exist in this lattice:
  compactons (ultra-localized waves with a compact support), usual solitary
  waves with exponential tails, as well as the corresponding kink-type
  solutions. These waves appear from rather general initial conditions and
  exist for long times. After very long transients they are destroyed due to
  inelastic collisions and evolve into phase chaos.}

Coupled autonomous oscillators are subject of high interest in nonlinear
science~\cite{Glass-01,Pikovsky-Rosenblum-Kurths-01}.  When the coupling of
the oscillators is weak, they can be described in the phase
approximation~\cite{Kuramoto-84}, where only a variation of the oscillator
phases matters.  The corresponding models are used for the description of
lattices~\cite{Ermentrout-Kopell-84,Kopell-Ermentrout-86,%
Sakaguchi-Shinomoto-Kuramoto-88,Ren-Ermentrout-00,Topaj-Pikovsky-02}, globally
coupled ensembles and
networks~\cite{Kuramoto-84,Kuramoto-75,Daido-92a,Daido-92,Strogatz-00}. In
the absence of coupling, the phase equations have only zero Lyapunov
exponents, therefore whether the phase dynamics is dissipative or conservative
depends solely on the properties of the coupling.  In studies which focus on
synchronization properties, one typically assumes that the coupling is
dissipative which thus tends to equalize the phases.  However, certain types
of coupling lead to a conservative dynamics -- a prominent example is a splay
state in a globally coupled ensemble of
oscillators~\cite{Tsang91,Nichols-Wiesenfeld-92,Strogatz-Mirollo-93,%
Watanabe-Strogatz-93,Watanabe-Strogatz-94}.

In the present work we consider the dynamics of a one-dimensional lattice of
oscillators with a dispersive, conservative coupling.  A realization of such a
lattice may be a multi-core fiber laser~\cite{Wrage_etal-00}, where individual
self-oscillating lasers are arranged in a ring, or an array of Josephson
junctions.  Since both, the local phase dynamics and the coupling are
non-dissipative, the dynamics is expected to be similar to that of the
well-known Hamiltonian type. An example of a Hamiltonian lattice is the
sine-Gordon lattice, for which the basic building blocks are traveling
solitary waves like pulses or kinks, that on integrable lattices collide
elastically, and in the non-integrable cases evolve into chaos~\cite{Scott-99}.

In recent years two new concepts appeared that have significantly extended our
understanding of nonlinear regimes in Hamiltonian lattices. One concept
introduces localized periodic breathers~\cite{Flach-Willis-98}; the other
introduces compact 
excitations in genuinely nonlinear lattices and wave
equations. Unlike the usual solitons that have exponential, or algebraic,
tails, the corresponding traveling waves have compact or almost compact
support. These waves, named \textbf{compactons}, have been introduced in
\cite{Rosenau-Hyman-93,Rosenau-94,Nesterenko-83}.

In the present paper we study general traveling waves in a chain of
dispersively coupled nonlinear self-sustained oscillators, continuing our
previous works~\cite{Rosenau-Pikovsky-05,Pikovsky-Rosenau-06} focused on
compacton solutions in phase oscillator chains.  Contrary to
\cite{Rosenau-Pikovsky-05,Pikovsky-Rosenau-06}, we consider here a generic
phase model without additional symmetries.  We show that dispersively coupled
oscillators possess not only compactons, but also ``classical'' solitary waves
with exponential tails. In dependence on parameters of the coupling and on the
velocity these solution bifurcate.  Remarkably, some of the compact solutions
described in \cite{Rosenau-Pikovsky-05,Pikovsky-Rosenau-06} exist only in the
symmetric situation, while other can be continued to the non-symmetric case.
In the general, non-symmetric case, we describe a novel type of semi-compact
waves which are compact from only one side and have an exponential tail on the
other side. Another novel feature appears in the analysis of traveling waves
on the lattice: here we describe a particular bifurcation from monotonic to
oscillatory tails of a solitary wave, such a bifurcation does not exist in the
quasicontinuous approximation where waves are described by means of a partial
differential equation. We also report on a remarkable property of the solitary
waves with exponential tails: they are extremely stable to collisions, so that
a chaotization in a finite lattice occurs -- contrary to the case of
compactons -- after exponentially long transients.

\section{\label{sec:model}Basic model}
We consider the phase dynamics of a chain of identical self-sustained
oscillators with frequency $\omega$
\begin{equation}
\frac{\de \varphi_n}{\de t} = \omega + q(\varphi_{n-1}-\varphi_n) +
q(\varphi_{n+1}-\varphi_n) 
\label{eq:phases} \text{.}
\end{equation}
Here $\varphi_n$ is the phase of the $n$-th oscillator and
$q(\varphi)=q(\varphi+2\pi)$ is the coupling function.  Hereafter, we assume
that this function is even, what corresponds to dispersive coupling, see
\cite{Pikovsky-Rosenau-06} for details. Introducing the phase differences
$v_n=\varphi_{n+1}-\varphi_n$ equation \eref{eq:phases} can be rewritten as
\begin{equation}
\frac{\de v_n}{\de t} = \nabla_d q(v) = q(v_{n+1}) - q(v_{n-1}) 
\label{eq:model} \text{.}
\end{equation}

The simplest possible even $2\pi$-periodic coupling function is $q(v)=\cos v$;
this case was studied in \cite{Pikovsky-Rosenau-06}, where it was shown, that
compactons and kovatons (glued together compact kink-antikink pairs) are
solutions arising out of the background $v^*=0$. Here, we extend the results
of \cite{Pikovsky-Rosenau-06} and demonstrate that this lattice also bears
periodic waves and solitons, as well as semi-compact waves. In
\cite{Pikovsky-Rosenau-06} it was also shown, that Eq.~\eref{eq:model} is a
Hamiltonian system. It possesses several integrals but is
non-integrable. Additional symmetries in $q(v)$ may result in additional
symmetries in Eq.~\eref{eq:model}, we will encounter them below.

Our main analytic tool for the study of the wave solutions of (\ref{eq:model})
is the quasi-continuous approximation (QCA). Associating $n$ with a continuous
variable $x$, one develops $q(v_{n\pm1})$ by the Taylor series expansion up to
the third order
\begin{equation}
q(v_{n\pm1}) = \left[ 1 \pm \frac{\partial}{\partial x} + \frac{1}{2}
\frac{\partial^2}{\partial x^2} \pm \frac{1}{6} \frac{\partial^3}{\partial
x^3} \right] q(v) \text{.}
\end{equation}
Inserting this into (\ref{eq:model}) results in
the partial differential equation
\begin{equation}
\frac{\partial v}{\partial t} = 2 \left[ 
\frac{\partial}{\partial x} + \frac{1}{6} \frac{\partial^3}{\partial x^3}
\right] q(v) 
\text{,} \label{eq:qca}
\end{equation}
which is the QCA for the lattice. We stress here, that this approximation
is not based on a small parameter: because spacing between the sites is one,
the higher-order terms are in general of the same order as the lower-order
ones. Thus the validity of this approximation may be supported only by a
comparison of its predictions (Section \ref{seq:qca}) with the numerical
solutions of the full equations (\ref{eq:model}) (Section \ref{sec:lattice}).

\section{Traveling waves in the quasi-continuum \label{seq:qca}}

In this section we analyze waves in the QCA~\eref{eq:qca}. First, we note that
any constant $v=v^*$ is a solution. We will look for waves on the base of such
a flat profile. Inserting the traveling wave ansatz $v(x,t)=v(x-\lambda
t)=v(s)$, where $\lambda$ is the wave velocity, into Eq.~\eref{eq:qca} and
integrating once yields
\begin{equation}
\lambda(v-v^*)+2(q(v)-q(v^*))+\frac{1}{3}\frac{\de^2}{\de s^2}q(v)=0
\text{.} \label{eq:qca_twa}
\end{equation}
The integration starts at $s_0$, where it is assumed, that
$v(s_0)=v^*=\text{const}$. This is not valid for periodic waves, where one has
to introduce the curvature of $q(v(s_0))$; however the resulting equations are
equivalent to those derived below. We multiply Eq.~\eref{eq:qca_twa} with
$\displaystyle{{\de q(v)}/{\de s}}$ and integrate again, to obtain
\begin{equation}
\lambda \Big[ q(v)\big(v-v^*\big)-Q(v,v^*) \Big] + \big( q(v)-q(v^*)\big)^2 
+ \frac{1}{6} \left[ \frac{\de q}{\de s} \right]^2 = 0
\text{.}
\label{eq:eqqs}
\end{equation}
The function $Q(v,v^*)$ is defined as
\begin{equation}
Q(v,u)=\int_{u}^v q(x) \de x \label{eq:Q}
\end{equation}
Equation \eref{eq:eqqs} can be rewritten as
\begin{equation}
\left( \frac{\de q}{\de v} \right)^2 \Bigg[ \frac{1}{2} \left( \frac{\de
v}{\de s} \right)^2 + U(v) \Bigg] = 0
\text{,} \label{eq:qca_pot}
\end{equation}
with the potential
\begin{equation}
U(v) = 3 \frac{ \lambda [ q(v) (v-v^*)-Q(v,v^*) ] + \big( q(v)-q(v^*)\big)^2
}{(q'(v))^2}
\text{.} \label{eq:qca_pot_func}
\end{equation}
Besides Eq.~\eref{eq:qca_pot} one can also derive a system of first-order ODEs
from Eq.~\eref{eq:qca_twa}, which reads
\begin{equation}
\frac{\de v}{\de s} = u \quad \text{,} \quad \frac{\de u}{\de s} = - \frac{
3\lambda (v-v^*) + 6 \big( q(v)-q(v^*) \big) + q''(v) u^2}{q'(v)} \text{.}
\label{eq:qca_ode}
\end{equation}

In the following we also need the properties of the linear approximation
$q(v)=q(v^*)+q'(v^*)(v-v^*)$, then the equations \eref{eq:qca_ode} simplify to
\begin{equation}
\frac{\de v}{\de s} = u \quad \text{,} \quad 
\frac{\de u}{\de s} = -\frac{3\lambda+6q'(v^*)}{q'(v^*)}(v-v^*)
\text{,} \label{eq:qca_ode_lin}
\end{equation}
provided that $q'(v^*)\neq0$. The stability of the 
fixed point at $v=v^*$ is determined by the eigenvalues of the
Jacobian 
\begin{equation}
l_{1,2}=\pm \sqrt{-(3\lambda+6q'(v^*))/q'(v^*)}.
\label{eq:qca_ode_l}
\end{equation}

\subsection{Solitary waves}

Solitary waves on the base of the constant field $v=v^*$ are the homoclinic
orbits of Eq.~\eref{eq:qca_pot} and Eq.~\eref{eq:qca_ode}. They start at
$v^*$, grow to a peak at $v_m$, and then go back to their origin $v^*$. An
equation for the wave velocity $\lambda_S$ can be obtained from the condition
$U(v_m)=0$ which immediately yields
\begin{equation}
\lambda_S =\frac{ \big(q(v_m)-q(v^*)\big)^2}{Q(v_m,v^*)-q(v_m)(v_m-v^*)}
\text{.} \label{eq:qca_sol_cond}
\end{equation}

\paragraph*{\bf Solitary waves with exponential tails.}
For the existence of a homoclinic trajectory one needs the fixed point of
Eq.~\eref{eq:qca_ode} to have one stable and one unstable direction, hence $v^*$
has to fulfill
\begin{equation}
-\frac{3 \lambda + 6 q'(v^*)}{q'(v^*)} > 0
\text{.} \label{eq:qca_sol_cond2}
\end{equation}
This condition yields a critical velocity $\lambda_C=-2 q'(v)$, which
separates a saddle-type stationary solution from a center.

A solitary wave with exponential tails is shown in
Fig.~\ref{fig:soliton_simple} below. The coupling function is $q(v)=\cos(v)$
and the background is $v^*=\pi /4$. The wave velocity is a free parameter, but
bounded by the above condition. For the case $v^*=\pi/4$ and $v_m > v^*$ this
results in $\lambda > 2 \sin \pi/4 = \sqrt{2}$. In
Fig.~\ref{fig:soliton_simple} the wave velocity was chosen to
$\lambda=\pi/2$. The tails of the solitary wave decay exponentially,
corresponding to the eigenvalues of the stationary state $v^*$.

\paragraph*{\bf Solitary waves with compact tails -- Compactons.}
Compactification may occur, if $q'(v^*)=0$. Linear waves do not exist around
this point and the approximation in Eq.~\eref{eq:qca_ode_lin} does not
hold. Therefore, one needs to approximate $q(v)$ to the second order in the
vicinity of $v^*$. When this approximation is inserted into Eq.~\eref{eq:qca}
one obtains the $K(2,2)$ equation, defined in \cite{Rosenau-Hyman-93}. For
$q(v)\approx q(v^*)+a v^2$ a solution of \eref{eq:qca_twa} is given by
\begin{equation}
v(s)+v^*=
\begin{cases} -\frac{2\lambda}{3a}\cos^2\left(\sqrt{\frac{3}{8}} s \right) &
|x|<\pi\sqrt{\frac{2}{3}} \\
0 & \text{else.} \end{cases} \label{eq:qca_twa_k22_sol}
\end{equation}
Usually, one cannot match two different solutions of one ODE, but here, the
highest order operator degenerates at $v=v^*$ and the solutions uniqueness is
lost. In the surrounding of $v^*$ the solution will behave like the $K(2,2)$
and compact waves occur in the full phase equation \eref{eq:qca}. In
Fig.~\ref{fig:compacton_simple} below we show a compacton for the coupling
function $q(v)=\cos v$ and $v^*=0$. The wave velocity was chosen to
$\lambda=2/\pi$.

\subsection{Kinks}

\begin{figure}
  \psfrag{vbar}[][][3]{$\bar{\text{v}}^*$}
  \psfrag{vo}[][][3]{$\text{v}^*$}
  \begin{center}
    \includegraphics[draft=false,width=0.5\textwidth]{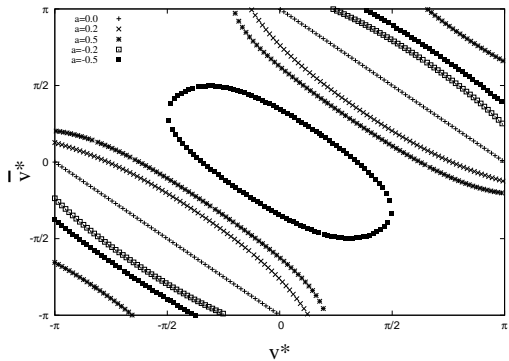}
  \end{center}
  \caption{The values of uniform background states that can be connected by 
  kinks, in the QCA for the particular
  coupling $q(v)=\cos v + a \cos 2 v$ and various values of $a$.}
  \label{fig:kink_points}
\end{figure}

The second class of traveling wave solutions are kinks, which correspond to
heteroclinic orbits between two uniform states $v^*$ and $\bar{v}^*$. The
height $\bar{v}^*$ of the kink has to fulfill condition \eref{eq:qca_sol_cond}
and furthermore it has to be a fixed point of \eref{eq:qca_pot}, meaning that
$U'(\bar{v}^*)=0$. This gives the following condition for the speed of the
kink (it can also be derived from \eref{eq:qca_twa} where one assumes a
constant solution $v_m=\text{const}$):
\begin{equation}
\lambda_K = 2 \frac{q(\bar{v}^*)-q(v^*)}{v^*-\bar{v}^*}
\text{.} \label{eq:qca_kink_cond}
\end{equation}
Combining the velocity relations \eref{eq:qca_sol_cond} and
\eref{eq:qca_kink_cond} by $\lambda_S=\lambda_K$ results in the final
condition for the height of the kink
\begin{equation}
Q(\bar{v}^*,v^*) = \frac{(q(\bar{v}^*)+q(v^*))(\bar{v}^*-v^*)}{2}
\text{.}
\label{eq:kink_height}
\end{equation}
In Fig.~\ref{fig:kink_points} the values of possible uniform states connected
by a kink are shown for the particular coupling function $q(v)=\cos v + a \cos
2 v$. Note, that a bifurcation occurs at $a=\pm 1/4$ and two new branches of
kink points emerge.
For $q(v)=\cos v$ one finds $\bar{v}^*=\pi-v^*$ and the velocity belonging to
this kink is $\lambda_{max}=\lambda_K=4 \cos v^*/(\pi-2 v^*)$.

\paragraph*{\bf Kinks with exponential tails.}
For kinks with exponential tails the same condition \eref{eq:qca_sol_cond2} as
for the solitary waves has to be fulfilled. In Fig.~\ref{fig:kink_simple} a
kink with exponential tails is shown. The coupling function is $q(v)=\cos(v)$
and $v^*=\pi/4$, $\bar{v}^*=3\pi/ 4$. The velocity is $\lambda =
\sqrt{32}/\pi$.

\paragraph*{\bf Compact kinks -- Kovatons.}
If $v^*$ and $\bar{v}^*$ fulfill the compactification condition
$q'(v^*)=q'(\bar{v}^*)=0$, both tails will become compact and may form a
compact kink-antikink pair, named {\it kovaton} \cite{Rosenau-Pikovsky-05}. An
example of this wave form is shown in Fig.~\ref{fig:kovaton_simple}, with
$q(v)=\cos v$ and $v^*=0$, $\bar{v}^*=\pi$ and $\lambda_K=4 / \pi$.

\paragraph*{\bf Exponential-compact kinks.}
In addition to kinks with exponential tails and kinks with compacton tails one
can also observe {\em semicompact} kinks with one exponential and one compact
tail. Consider the coupling $q(v)=\cos v + a \cos 2v$ with $a=0.2$. In this
special setup $v^*=0$ fulfills the compactification condition and $\bar{v}^* =
2.39955$ is the kink point satisfying \eref{eq:kink_height} with velocity
\eref{eq:qca_kink_cond} $\lambda_K = 1.60011$. This kink is shown in
Fig.~\ref{fig:comp_exp_kink}. It is compact at $v=v^*$ and exponential at
$v=\bar{v}^*$.

\subsection{Periodic waves}

Periodic waves around $v^*$ exist if the eigenvalues $l_{1,2}$ in
\eref{eq:qca_ode_l} are purely imaginary. A periodic wave in the QCA is shown
in Fig.~\ref{fig:periodic_wave_simple}.  The velocity of a periodic wave must
satisfy the condition resulting from \eref{eq:qca_ode}:
\begin{equation}
-\frac{3 \lambda + 6 q'(v^*)}{q'(v^*)} < 0
\text{.} \label{eq:qca_periodic_cond}
\end{equation}
For $q(v)=\cos v$ and $0< v^*<\pi$ this condition yields $\lambda<2 \sin
v^*$. At $\lambda=\lambda_C=-2 q'(v^*)$ some kind of bifurcation occurs.

\begin{figure}
  \begin{center}
    \subfigure[$\lambda=\lambda_{S,min}-0.1$]{\label{fig:phase_space_a}\includegraphics[draft=false,angle=270,width=0.4\textwidth]{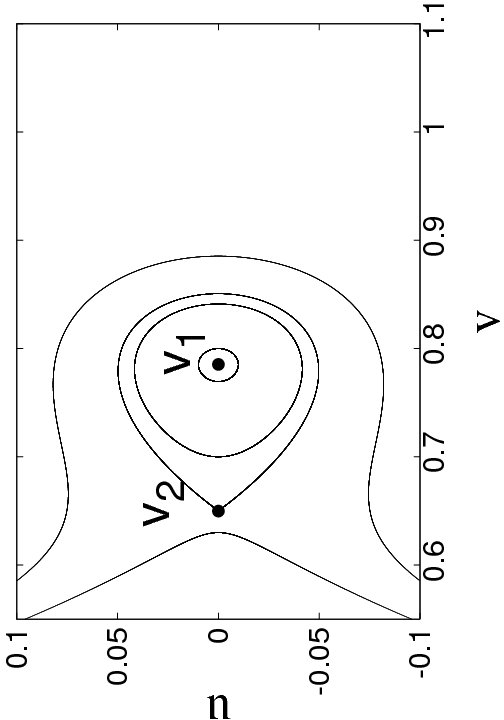}}
    \subfigure[$\lambda=\lambda_{S,min}+0.1$]{\label{fig:phase_space_b}\includegraphics[draft=false,angle=270,width=0.4\textwidth]{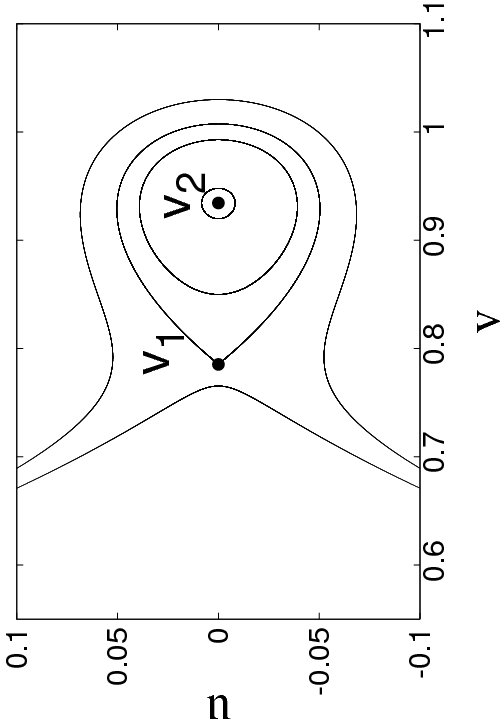}}
  \end{center}
  \caption{Phase space of \eref{eq:qca_ode} for $v^*=\pi/4$ and a)
  $\lambda=\lambda_C-0.1$ and b) $\lambda=\lambda_C+0.1$. The critical
  velocity is $\lambda_C=2\sin \pi/4 = \sqrt{2}$. It is clearly visible that
  below the critical velocity the fixed point $(v_1,u_1)$ is a center, hence
  periodic waves exist in \eref{eq:qca}. Above $\lambda_C$ one can observe
  solitonic solutions: the background $v_1$ is a saddle point and a homoclinic
  orbit exists.}
  \label{fig:phase_space}
\end{figure}

To quantify the dynamical behavior of the QCA near $\lambda_C$ we
simplify \eref{eq:qca_ode} to
\begin{equation}
\frac{\de v}{\de s} = u \quad \text{,} \quad \frac{\de u}{\de s} = \lambda
(v-v^*) - v^2-{v^*}^2 \text{.} \label{eq:qca_ode_quadratic}
\end{equation}
This ODE is not an approximation of \eref{eq:qca_ode} in a strict sense, since
we have neglected the terms $q''(v)u^2$ and $q'(v)$, but the qualitative
behavior does not change. System \eref{eq:qca_ode_quadratic} has two fixed
points $(v_1,u_1)=(v^*,0)$ and $(v_2,u_2)=(\lambda-v^*,0)$. The eigenvalues of
the corresponding Jacobian are
\[
l_{1,2}^{(1)}  =  \pm \sqrt{\lambda-2v^*} \;\qquad
l_{1,2}^{(2)}  =  \pm \sqrt{2v^*-\lambda} \text{.}
\]
If the velocity reaches the critical value $\lambda_C=2v^*$, the two fixed
points coincide and furthermore both fixed points change their roles; if
$\lambda<\lambda_C$ the center is $(v_1,u_1)$ and $(v_2,u_2)$ is a
saddle-point; if $\lambda>\lambda_C$ the roles have changed and $(v_1,u_1)$ is
the saddle and $(v_2,u_2)$ is the center. Now we return to
Eq.~\eref{eq:qca_ode}. In Fig.~\ref{fig:phase_space} the phase space of
\eref{eq:qca_ode} near the critical velocity is shown. If $\lambda<\lambda_C$
the first fixed point will be a center, hence periodic waves exists in the QCA
and for $\lambda>\lambda_C$ the fixed point is a saddle point. Furthermore,
the unstable and the stable manifold of this saddle are connected -- a
homoclinic orbit exists, resulting in solitonic solutions in the QCA. So, at
the critical velocity we have a transition or a bifurcation from periodic
waves to solitons.

%
%\paragraph*{Remark:}
%It should be noted, that any point with $q'(v^*)=0$ can only be crossed by a
%trajectory, if the velocity takes one specific value, such that the numerators
%of Eq.~\eref{eq:qca_ode_func} and Eq.~\eref{eq:qca_pot_func} vanish to zero
%and the values of \eref{eq:qca_ode_func} and \eref{eq:qca_pot_func} are
%finite, hence \eref{eq:qca_kink_cond} is fulfilled for $v_c$. The number of
%possible velocities is finite for an given $v^*$. So , one can say that this
%effect separates the phase space at $v_c$.

%\paragraph*{Questions:} 
%\begin{itemize}
%\item What happens for $q'=q''=0$?
%\end{itemize}

\section{\label{sec:lattice}Traveling waves in the lattice}

Now we turn our attention to traveling waves in the full phase lattice model
\eref{eq:model}. The wave ansatz for this model can be formulated as
\begin{equation}
v_n(t)=v(n-\lambda t)=v(s)
 \label{eq:lattice_twa_ansatz}
\end{equation}
with velocity $\lambda$. Inserting \eref{eq:lattice_twa_ansatz} into
\eref{eq:model} yields
\begin{equation}
\dot{v}=\frac{1}{\lambda} \Big(q\big(v(s-1)\big)-q\big(v(s+1)\big) \Big)
\text{.} \label{eq:lattice_twa}
\end{equation}
We integrate this equation from $s_0$ to $s$ to obtain 
\begin{equation}
v(s)-v^* = \frac{1}{\lambda} \int_{s-1}^{s+1} \big[q(v^*)-q(v(\tau))\big]\de
\tau
\label{eq:lattice_twa_int} \text{,}
\end{equation}
where it is supposed, that $v(s<s_0)=v^*$. Again, as in the continuous version,
the exact initial conditions $v(s_0)$ are not relevant, they can be absorbed
into the constant of integration. If one requires that $v(s)=\bar{v}^*$ is a
constant solution, hence a kink exist, the corresponding velocity $\lambda_K$
has to satisfy
\begin{equation}
\lambda_K=2 \frac{q(v^*)-q(\bar{v}^*)}{\bar{v}^*-v^*}
\text{.} \label{eq:lattice_cond}
\end{equation}
This condition is exactly analogous to condition \eref{eq:qca_kink_cond} in
the QCA.

\subsection{Fixed point analysis}

Similar to the fixed point analysis in the QCA, one can analyze the behavior
of traveling waves close to the background $v^*$. To this end we linearize
$q(v) \approx q(v^*)+a (v-v^*)$ in \eref{eq:lattice_twa} and apply the
exponential ansatz $v(t)=A \exp l t$. This yields the characteristic equation
\begin{equation}
l = \frac{a}{\lambda}( e^{-l} - e^l )
\text{.}
\end{equation}
Note again, that $a=q'(v^*)\neq 0$, meaning that this approximation does not
hold for the compacton backgrounds.

\begin{figure}
  \begin{center}
    (a) \includegraphics[draft=false,width=0.45\textwidth]{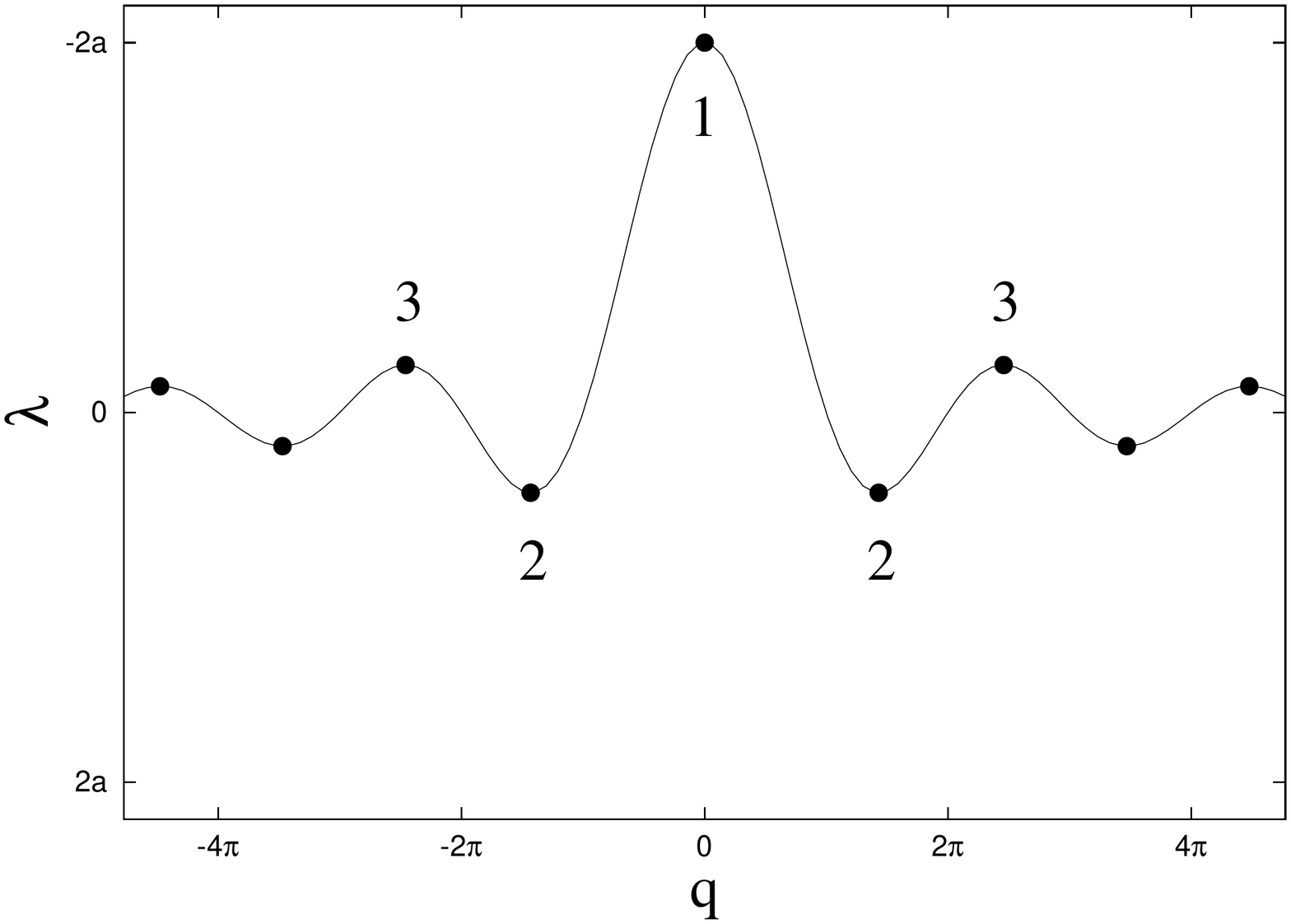}\hfill
(b) \includegraphics[draft=false,width=0.45\textwidth]{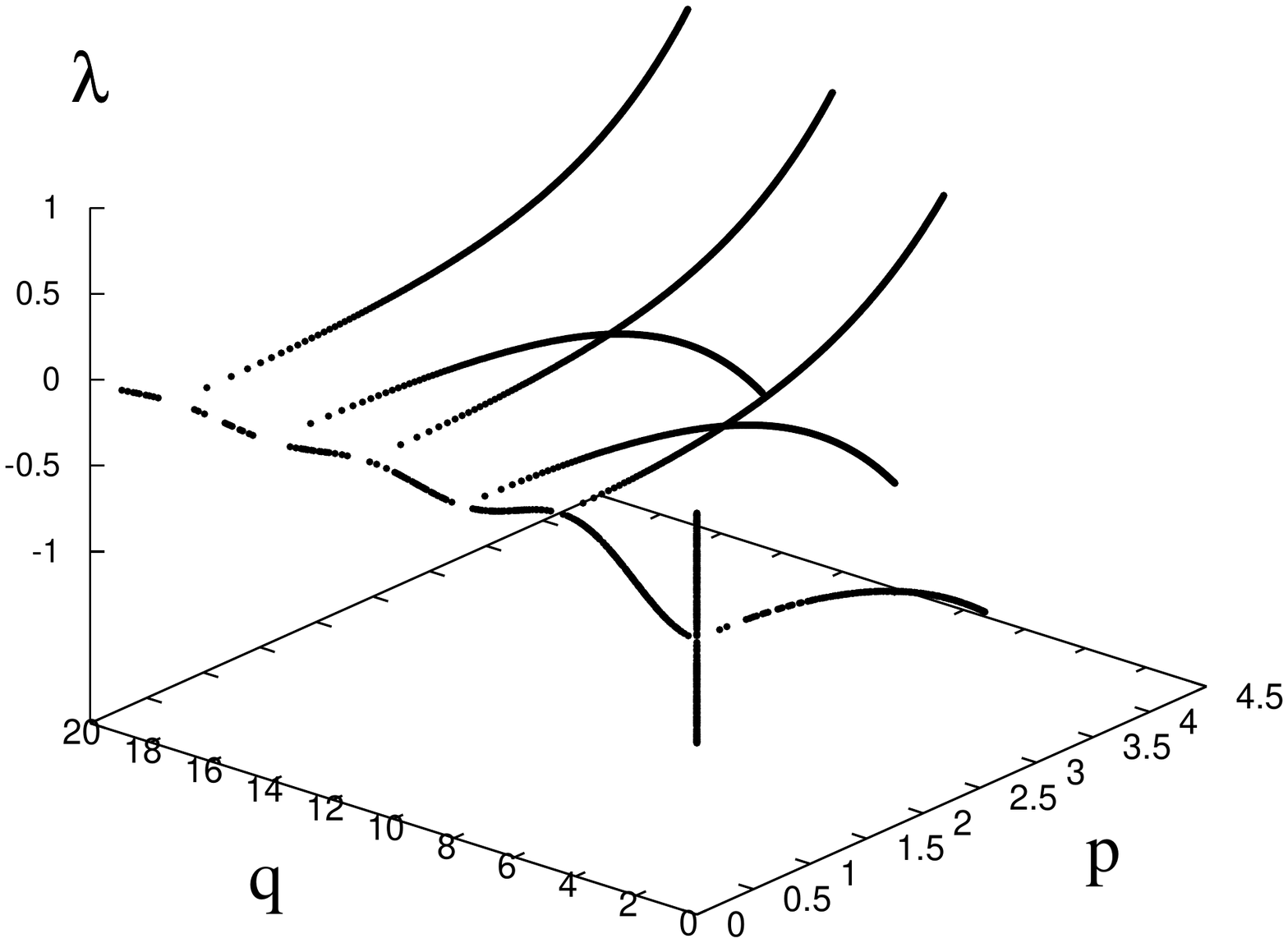}
  \end{center}
  \caption{(a): Wave speed $\lambda$ in dependence of the imaginary eigenvalue
  $q$, see \eref{eq:ev_im}. The dots mark possible bifurcation points to real
  eigenvalues. (b): Wave speed $\lambda$ vs. the real and imaginary part
  of the eigenvalues, with $a=\sin(0.2)$. Only the positive parts of the real
  and imaginary axes are shown.}
  \label{fig:ev}
\end{figure}

We split $l$ into its real and imaginary part $l=p+\ii q$ to obtain
\begin{equation}
\label{eq:eigenvalues}
p = - 2 \frac{a}{\lambda} \cos q \sinh p \qquad \text{and} \quad q = 2
\frac{a}{\lambda} \sin q \cosh p \text{.}
\end{equation}
For a purely imaginary eigenvalues $p=0$ we obtain
\begin{equation}
q= - 2 \frac{a}{\lambda} \sin q \quad \text{or} \quad 
\lambda= -2 a \frac{\sin q}{q} \text{.}
\label{eq:ev_im}
\end{equation}
This function is plotted in Fig.~\ref{fig:ev}(a). In this plot, the dots mark
possible points for transitions to eigenvalues with real parts. In
Fig.~\ref{fig:ev}(b) all eigenvalues $l=p+\ii q$ are shown. Purely real
eigenvalues are
\begin{equation}
p=-2\frac{a}{\lambda} \sinh p \quad \text{or} \quad
\lambda= -2 a \frac{\sinh p}{p} \text{.}
\label{eq:ev_re}
\end{equation}
So, when $\lambda$ crosses $-2a$ (point 1 in Fig.~\ref{fig:ev}(a)), a
bifurcation from two purely imaginary eigenvalues to two purely real
eigenvalues occurs. This scenario corresponds to the transition from periodic
to solitary waves and the critical velocity is $\lambda_C=-2 a$. The situation
is analogous to the bifurcation in the QCA and the critical velocity is the 
in both situations.

The next bifurcation occurs, when $\lambda$ crosses point 2, see
Fig.~\ref{fig:ev}. Then, a bifurcation from a center to a stable and an
unstable spiral point occurs. This refers to the transition from periodic
waves to solitary waves with oscillating tails and exponentially decaying
amplitude. Since the bifurcation occurs on the imaginary axis, one can
calculate the critical velocity from \eref{eq:eigenvalues} by setting
$\lambda'(q)=0$ and for the special case $q(v)=\cos v$ one obtains $\lambda
\approx -2 a \cdot 0.217$. Note, that there is no counterpart in the QCA for
this bifurcation.

\subsection{Numerical determination of traveling waves} 

From \eref{eq:lattice_twa_int} it is possible to construct a numerical scheme
to find solitary wave solutions of the lattice
\cite{Pikovsky-Rosenau-06,Petviashvili-76,Petviashvili-81} . One initially
guesses a wave profile $v_0(t)$ and than iterates
\begin{equation}
\label{eq:lattice_twa_scheme}
\tilde{v}(s)= v^*+\frac{1}{\lambda} \int_{s-1}^{s+1}\big(q(v^*)-
q(v_k(\tau))\big) \de \tau
\qquad
v_{k+1}=\left( \frac{||v_k||}{||\tilde{v}||} \right) ^{3/2} \tilde{v}
\text{,} 
\end{equation}
$||\cdot||$ denotes the $L_1$-norm. The integral is calculated 
by a high order
Lagrangian integration rule \cite{Abramowitz-Stegun-64}. To construct kink
solutions, one has to omit the normalization in \eref{eq:lattice_twa_scheme}
by setting $v_{k+1}=\tilde{v}$. Periodic waves can be obtained by a slight
modification of \eref{eq:lattice_twa_scheme}. Here, the wave length $w$ is
introduced and periodic boundary conditions $\tilde{v}(0)=\tilde{v}(w)$ are
assumed in \eref{eq:lattice_twa_scheme}.

We want to point out two issues one has to keep in mind when using this
algorithm. First, in \eref{eq:lattice_twa_scheme} the normalization exponent
$3/2$ is used. In a few cases this exponent is to large and has to be set to
smaller values, otherwise the algorithm will diverge. Secondly, for
backgrounds different from $0$ one has to shift the coordinates $v \mapsto v^*
+ v$.

\subsection{Solitary waves}

Solitary waves arise out of a background with one stable and one unstable
direction. So, they have to fulfill $(\lambda+2 q'(v^*))/q'(v^*)>0$ in order
to obtain two real eigenvalues of the fixed point.

\begin{figure}
  \begin{center}
    \includegraphics[draft=false,width=0.5\textwidth]{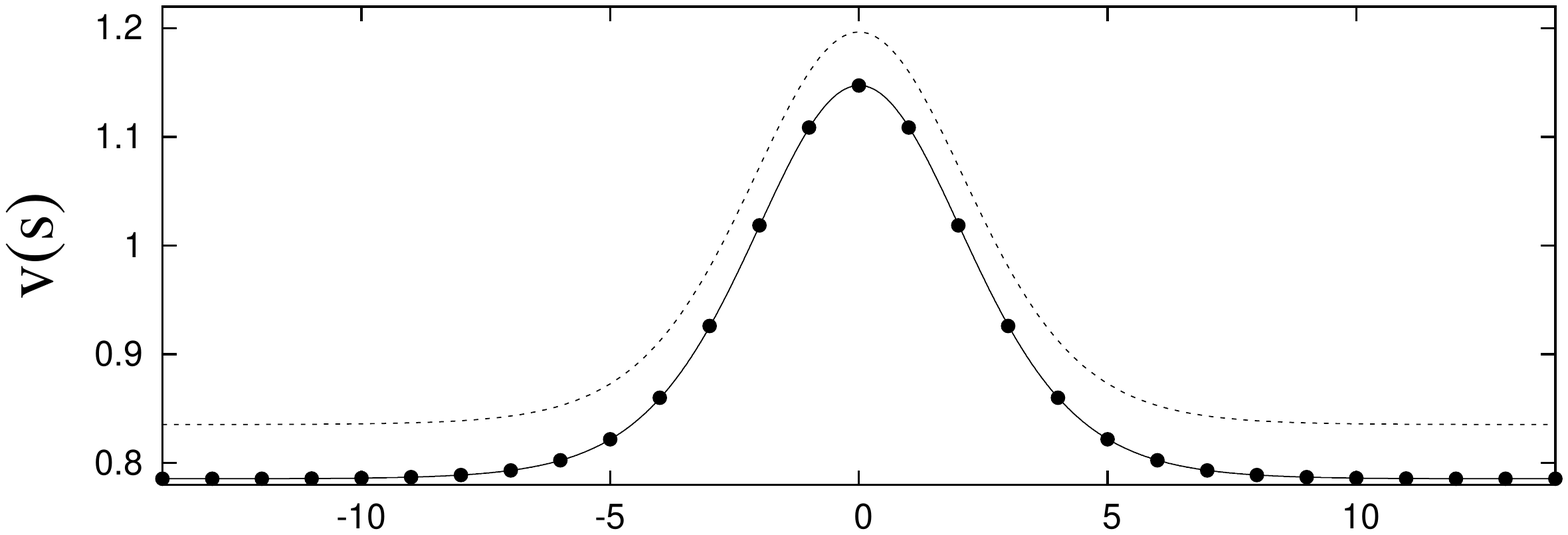}
    \includegraphics[draft=false,width=0.5\textwidth]{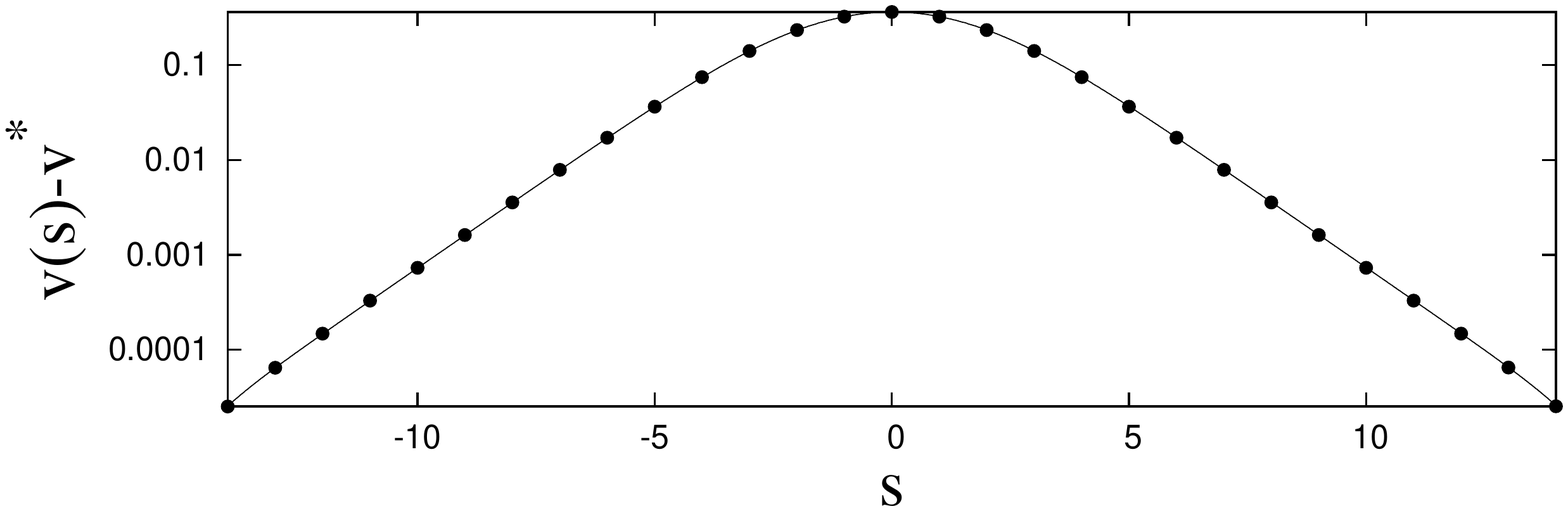}
  \end{center}
  \caption{Top panel: Soliton solution for the coupling $q(v)=\cos v$, which
  arise out of the background $v^*=\pi/4$. The form of the soliton was
  calculated numerically with the help of \eref{eq:lattice_twa_scheme} and the
  velocity is $\lambda=\pi/2$. Here and in the following figures bold 
  dots show the soliton on the lattice and
  the dashed line the solution of QCA \eref{eq:qca_twa} which has 
  an additional offset
  for better visibility. Bottom panel: the soliton in logarithmic scale.}
  \label{fig:soliton_simple}
\end{figure}

\paragraph*{\bf Solitary waves with exponential tails.}
In Fig.~\ref{fig:soliton_simple} we show a solitary wave with exponential
tails. The coupling is $q(v)=\cos v$ and the background is $v^*=\pi/4$. The
velocity of was chosen to $\lambda=\pi/2$, fulfilling the condition
$(\lambda+2 q'(v^*))/q'(v^*)>0$. The solitary wave was computed with the
scheme \eref{eq:lattice_twa_scheme}.

\begin{figure}
  \begin{center}
    \includegraphics[draft=false,width=0.5\textwidth]{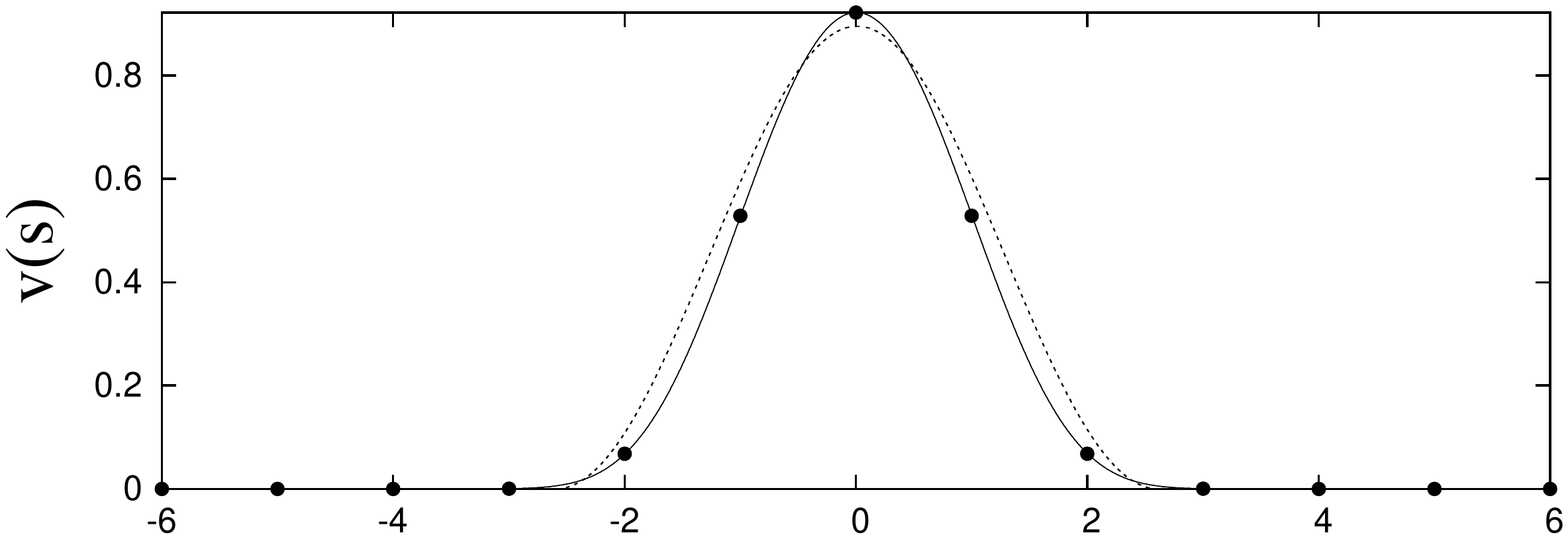}
    \includegraphics[draft=false,width=0.5\textwidth]{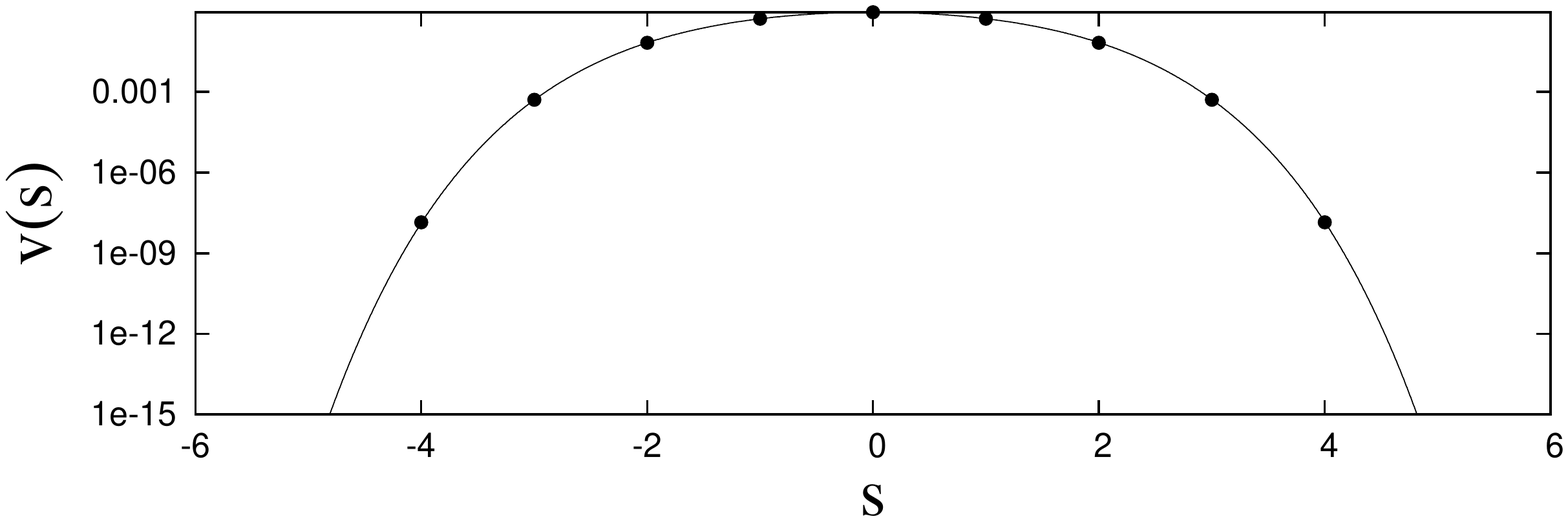}
  \end{center}
  \caption{Top panel: the shape of the compacton for the coupling $q(v)=\cos
  v$. The background is $v^*=0$ and the velocity was set to
  $\lambda=2/\pi$. Markers show the compacton on the discrete lattice and the
  dashed line is the corresponding solution in the quasi-continuum
  approximation. Bottom panel: the same plot in logarithmic scale. The super
  exponentially decaying tails are clearly visible.}
  \label{fig:compacton_simple}
\end{figure}

\paragraph*{\bf Compact solitary waves.}
Again, as in the QCA, the condition $q'(v^*)=0$ allows the compactification of
the tails. In Fig.~\ref{fig:compacton_simple} we show a compacton arising out
of the background $v^*=0$ for the coupling $q(v)=\cos v$. The wave velocity
was set here to $\lambda=2/\pi$. The compacton is not purely compact, but has
super-exponentially decaying tails \cite{Pikovsky-Rosenau-06}. Thus, although
there is a qualitative difference between the lattice and the QCA,
quantitatively these solutions are very close to each other.

\subsection{Kinks}

To derive an analogon to \eref{eq:kink_height}, we rewrite
Eq.~\eref{eq:lattice_cond} as $\lambda_K\big(\bar{v}^*-v^*\big)=2\big(
q(v^*)-q(\bar{v}^*) \big)$, multiply with $q'(\bar{v}^*)$, and integrate over
$\bar{v}^*$ from $v^*$ to $\bar{v}^*$ to obtain
\begin{equation}
\lambda_K \int_{v^*}^{\bar{v}^*} q'(v) \big( v - v^* ) \de v = 2
\int_{v^*}^{\bar{v}^*} q'(v) \Big( q(v^*)-q(v) \Big) \de v \text{,}
\end{equation}
and finally
\begin{equation}
\lambda_K = \frac{\big( q(v^*) - q(\bar{v}^*) \big)^2}
{Q(\bar{v}^*,v^*) - q(\bar{v}^*) \big (\bar{v}^*-v^* \big) }
\text{.} \label{eq:lattice_cond2}
\end{equation}
Combining Eq.~\eref{eq:lattice_cond} and \eref{eq:lattice_cond2} yields
\begin{equation}
Q(\bar{v}^*,v^*) = \frac{(q(\bar{v}^*)+q(v^*))(\bar{v}^*-v^*)}{2}
\text{,} \label{eq:kink_height2}
\end{equation}
which matches exactly the kink condition for the QCA.

\begin{figure}
  \begin{center}
    \includegraphics[draft=false,width=0.5\textwidth]{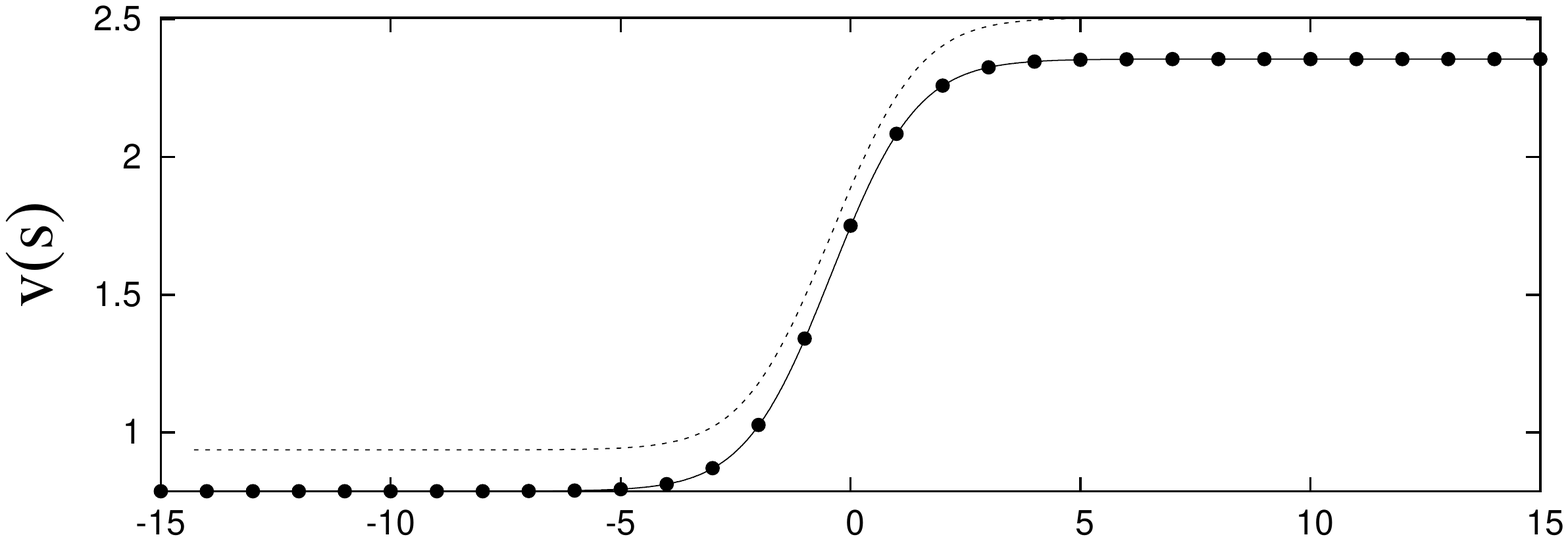}
    \includegraphics[draft=false,width=0.5\textwidth]{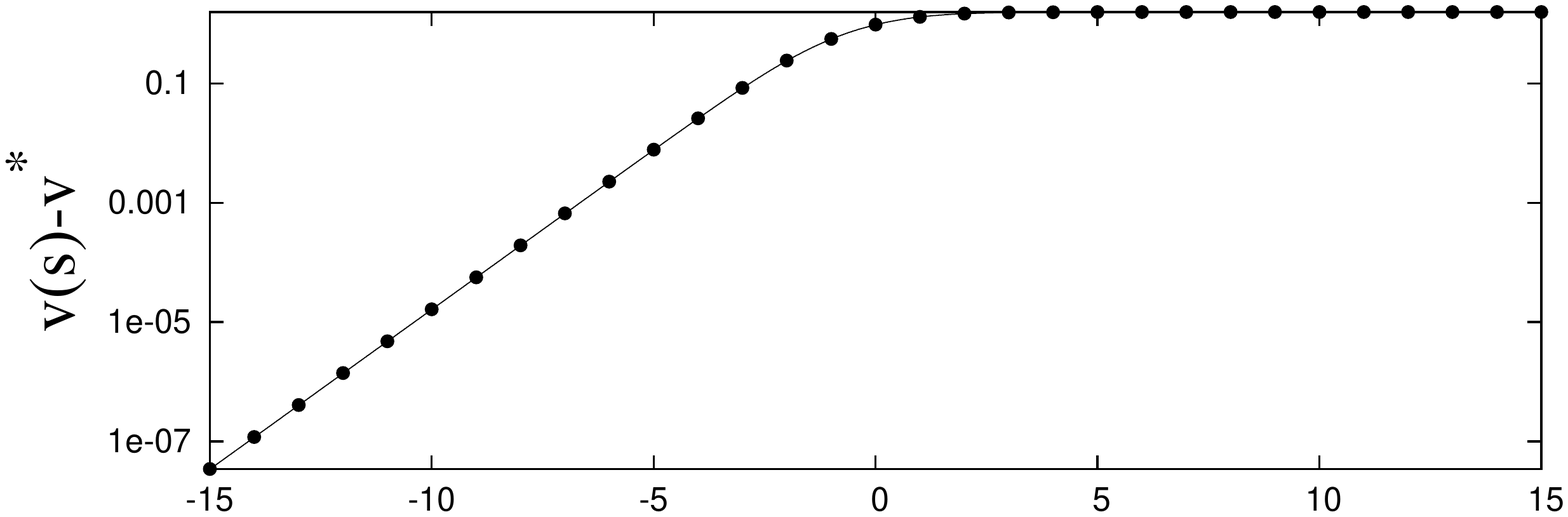}
    \includegraphics[draft=false,width=0.5\textwidth]{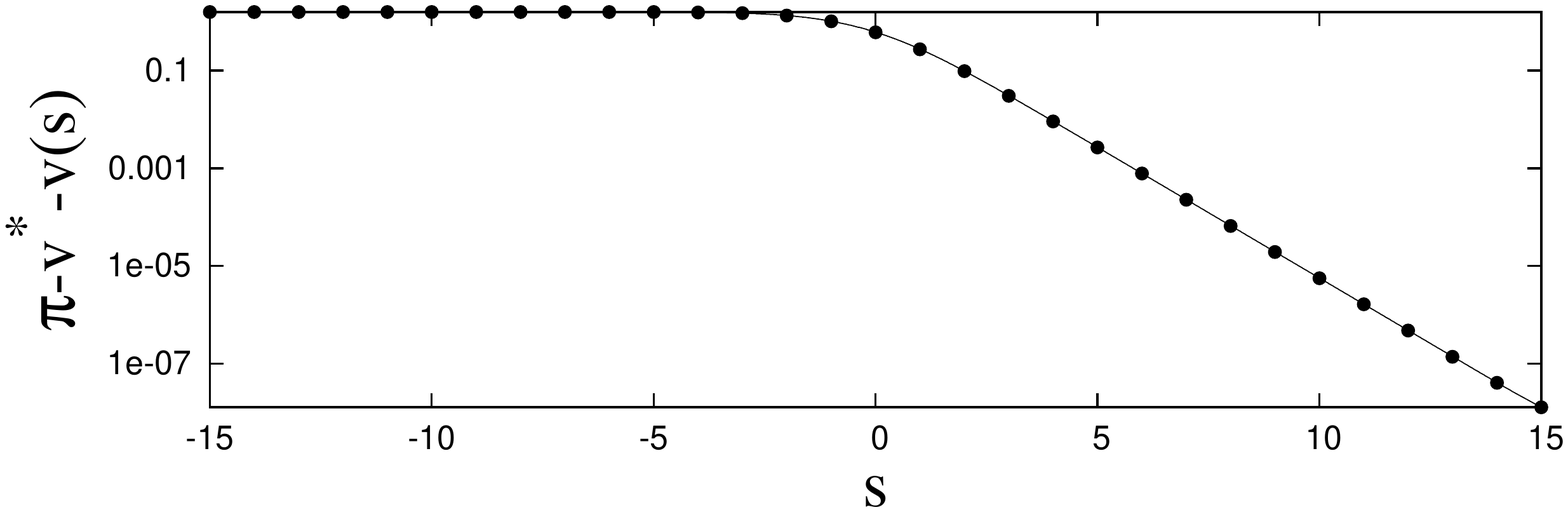}
  \end{center}
  \caption{Top panel: The shape of the kink for $v^*=\pi/4$, the velocity is
  $\lambda=\sqrt{32}/\pi$. Center panel: The kink in logarithmic scale. Bottom
  panel: The kink shown from its top in logarithmic scale. Markers show the
  kink on the lattice and the dashed line is the QCA with an additional offset
  for better visibility.}
  \label{fig:kink_simple}
\end{figure}

\paragraph*{\bf Kinks with exponential tails.}
In Fig.~\ref{fig:kink_simple} we show the shape of a kink. The coupling is
$q(v)=\cos v$ and the background is $v^*=\pi/4$. The velocity of the kink is
given by \eref{eq:lattice_cond} $\lambda_K=\sqrt{32}/\pi$.

\begin{figure}
  \begin{center}
    \includegraphics[draft=false,width=0.5\textwidth]{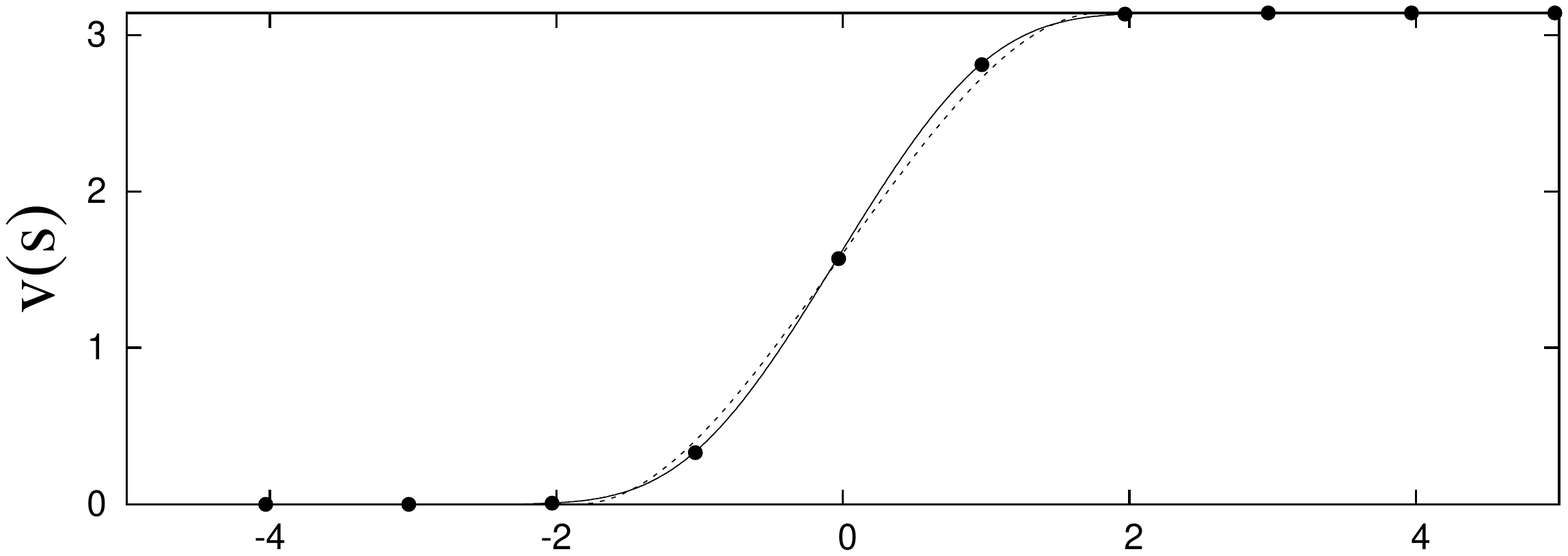}
    \includegraphics[draft=false,width=0.5\textwidth]{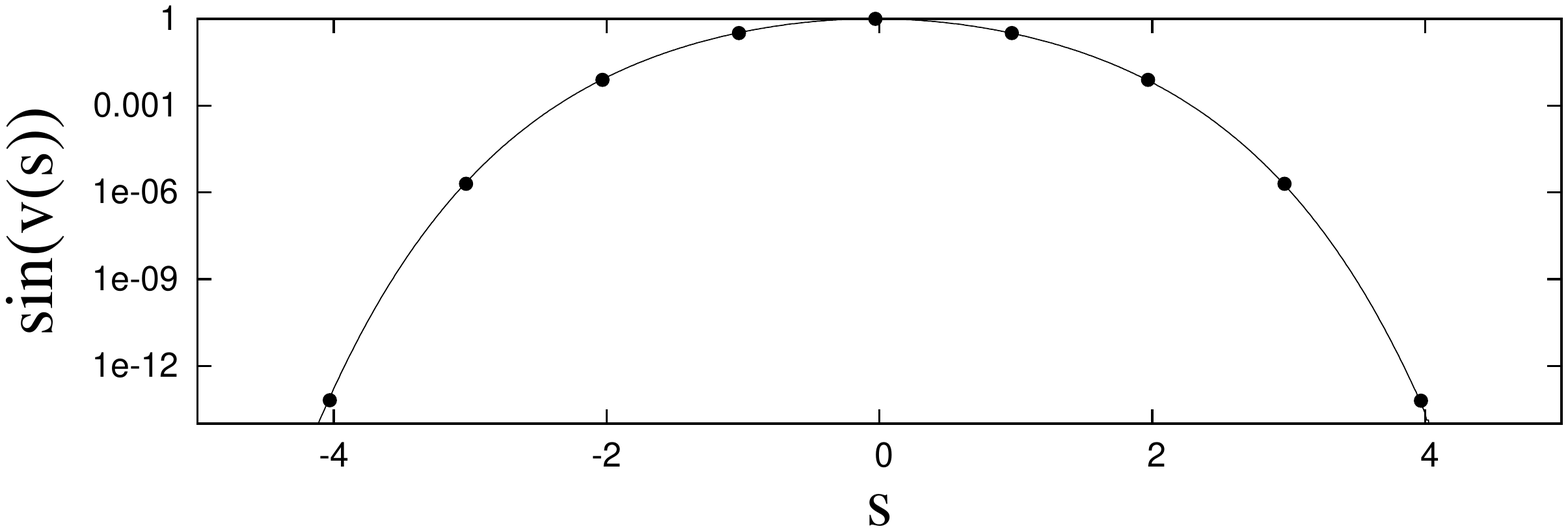}
  \end{center}
  \caption{Top panel: The wave form of the compact kink solution between
  $v^*=0$ and $\bar{v}^*=\pi$ for the coupling $q(v)=\cos v$. Markers show the
  wave form on the discrete lattice and the dashed line represents the
  quasi-continuum approximation of the kink. Bottom panel: The sine of the
  kink in logarithmic scale.}
  \label{fig:kovaton_simple}
\end{figure}

\paragraph*{\bf Kinks with compact tails.}
One can observe compact kinks, if $q'(v^*)=0$ and $q'(\bar{v}^*)=0$. For the
coupling $q(v)=\cos v$ such a case exists with $v^*=0$ and
$\bar{v}^*=\pi$. The shape of this compact kink is shown in
Fig.~\ref{fig:kovaton_simple}. Here, the velocity is $\lambda=4/\pi$.

\begin{figure}
  \begin{center}
    \includegraphics[draft=false,width=0.5\textwidth]{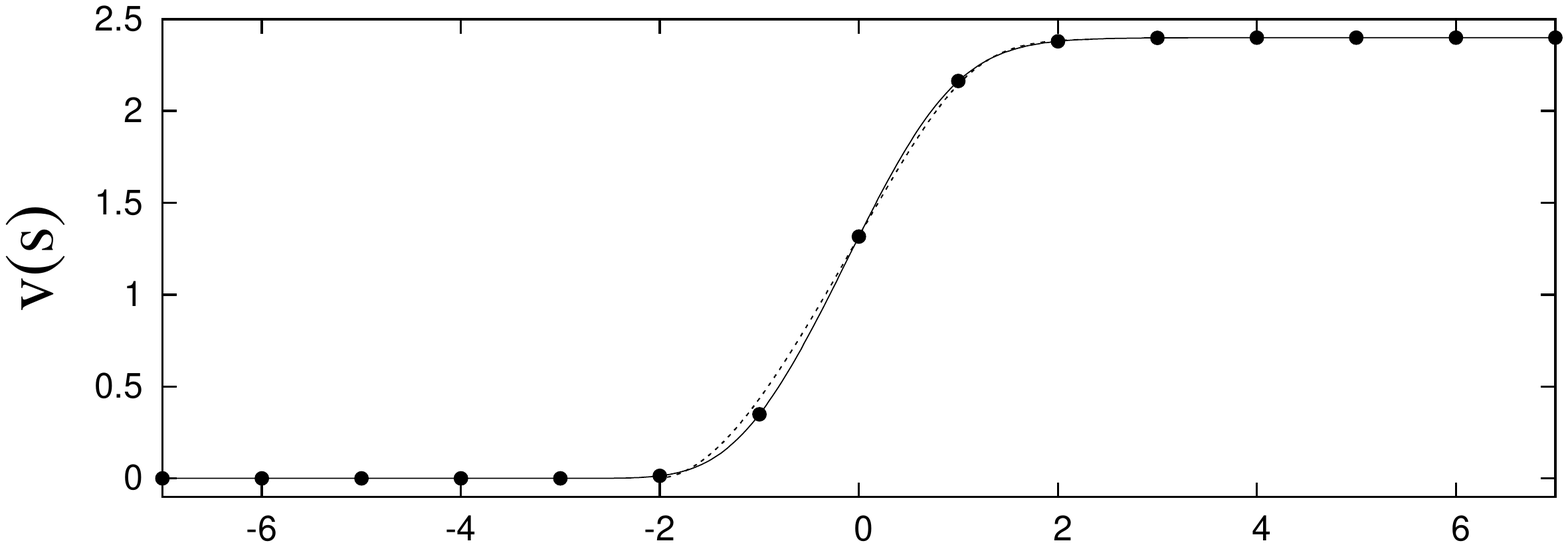}
    \includegraphics[draft=false,width=0.5\textwidth]{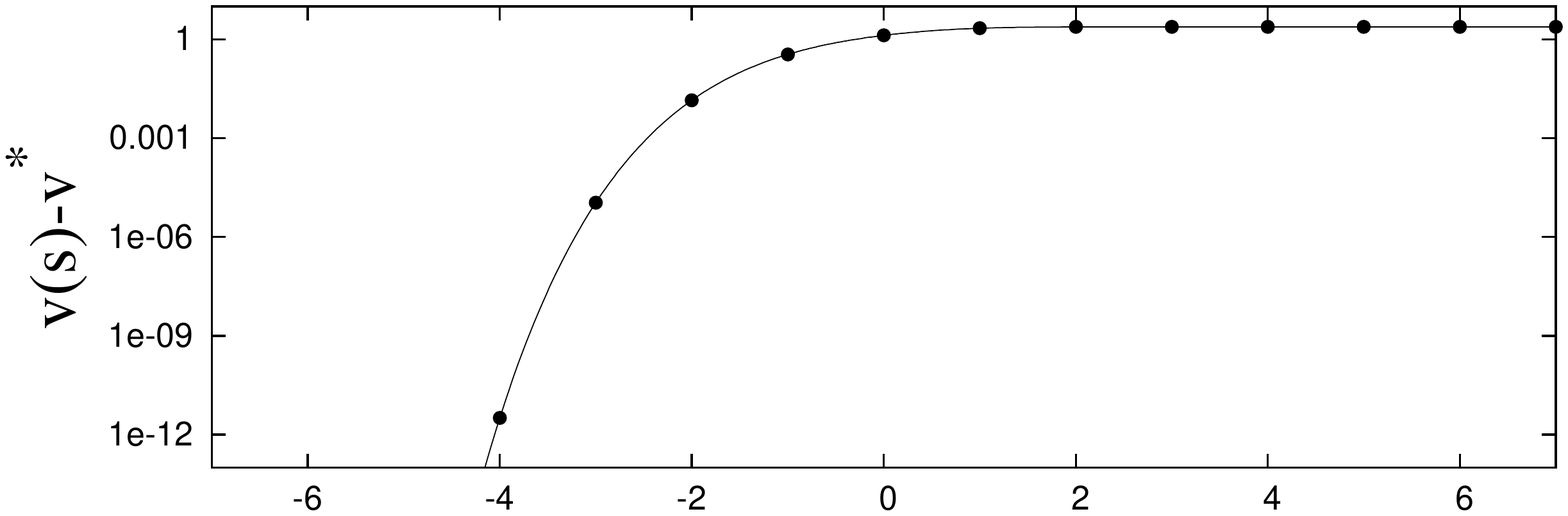}
    \includegraphics[draft=false,width=0.5\textwidth]{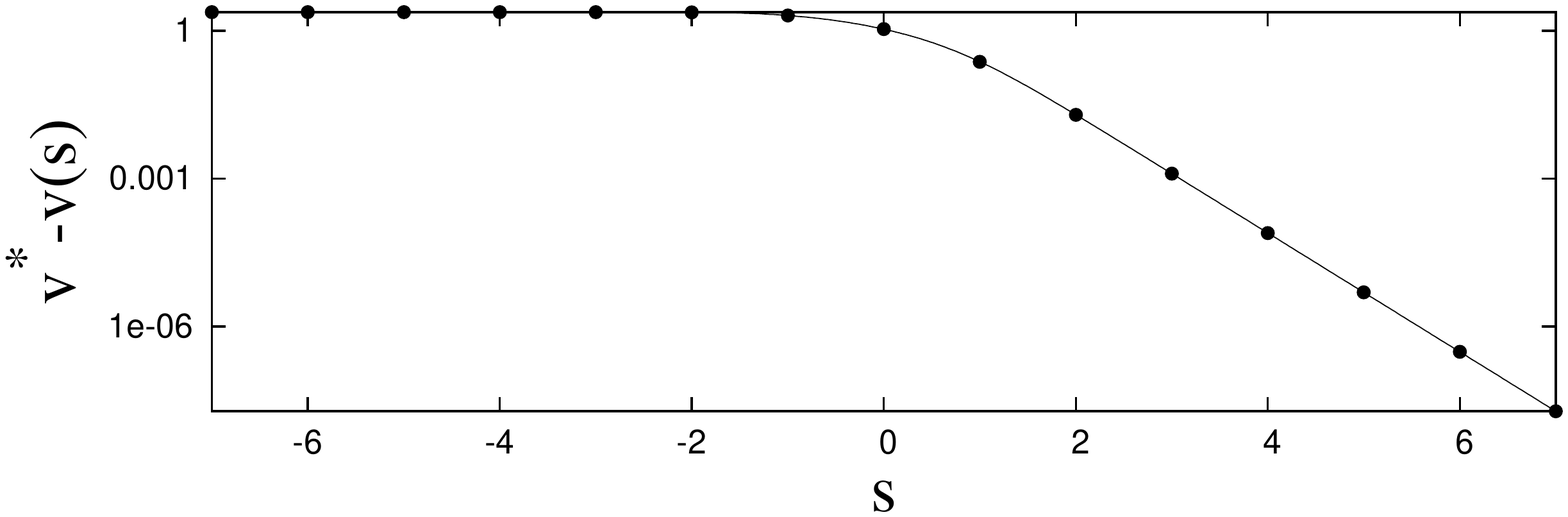}
  \end{center}
  \caption{Top panel: The shape of a kink with one exponential and one compact
  tail. The coupling for this specific wave is $q(v)=\cos v + a\cos 2v$ with
  $a=0.2$. The position of the kink point and the wave velocity can be
  obtained from \eref{eq:kink_height2} and \eref{eq:lattice_cond}, their
  values are $\lambda=1.60011$ and $\bar{v}^*=2.39955$. Middle panel: The kink
  in logarithmic scale. The compact tail of the kink is hear clearly
  visible. Bottom panel: The kink shown from its top in logarithmic scale, the
  exponential tail becomes visible.}
  \label{fig:comp_exp_kink}
\end{figure}

\paragraph*{\bf Semi-compact kinks.}
It also possible to observe kinks with one exponential decaying tail and one
compact tail. This is the case for $q(v)=\cos v + a \cos 2v$ with $a=0.2$. For
$\lambda=1.60011$ and $\bar{v}^*=2.39955$ the kink condition
\eref{eq:kink_height2} is satisfied and such a kink is found by the numerical
method described above, see Fig.~\ref{fig:comp_exp_kink}.

\subsection{Periodic waves}

\begin{figure}
  \begin{center}
    \includegraphics[draft=false,width=0.5\textwidth]{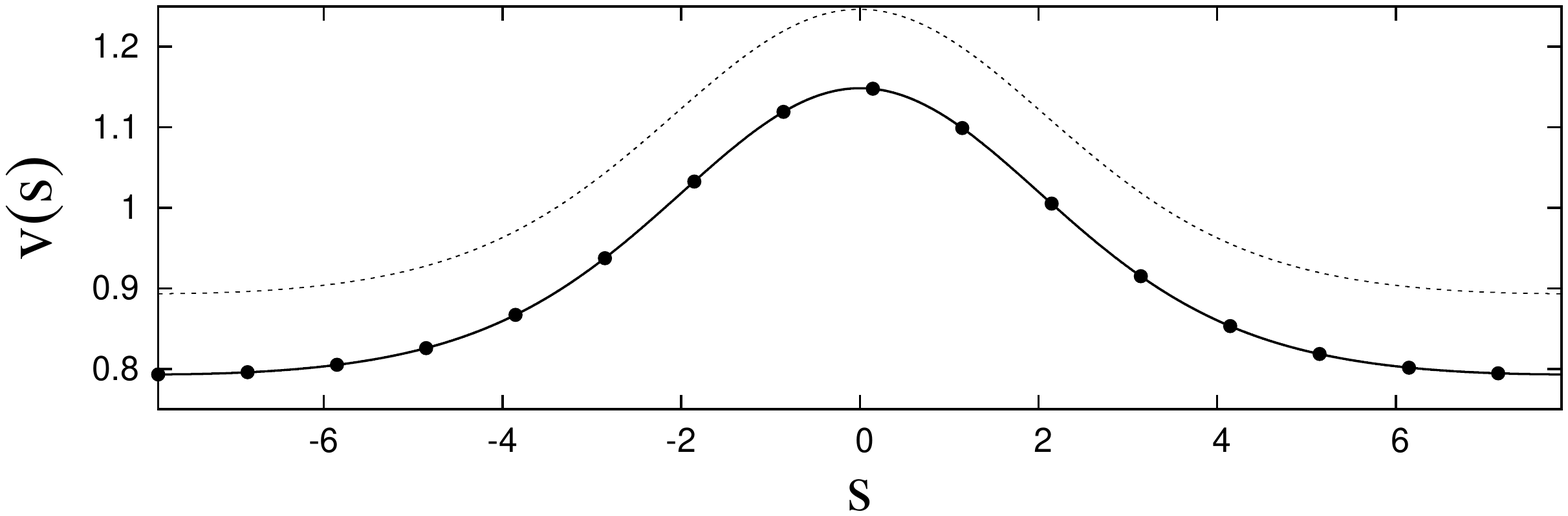}
  \end{center}
\caption{The shape of a periodic wave on the lattice. The offset is
  $v^*=\pi/4$, the wave length is $w=5\pi$ and the velocity $\lambda=\pi/2$.}
\label{fig:periodic_wave_simple}
\end{figure}

Periodic waves can be calculated with \eref{eq:lattice_twa_scheme} and
periodic boundary conditions. An example is shown in
Fig.~\ref{fig:periodic_wave_simple}. Here, the offset is $v^*=\pi/4$, the wave
length is $w=5 \pi$ and the velocity is $\lambda=\pi/2$.

\subsection{Solitary waves with periodically decaying tails}

\begin{figure}
  \begin{center}
    \includegraphics[draft=false,width=0.5\textwidth]{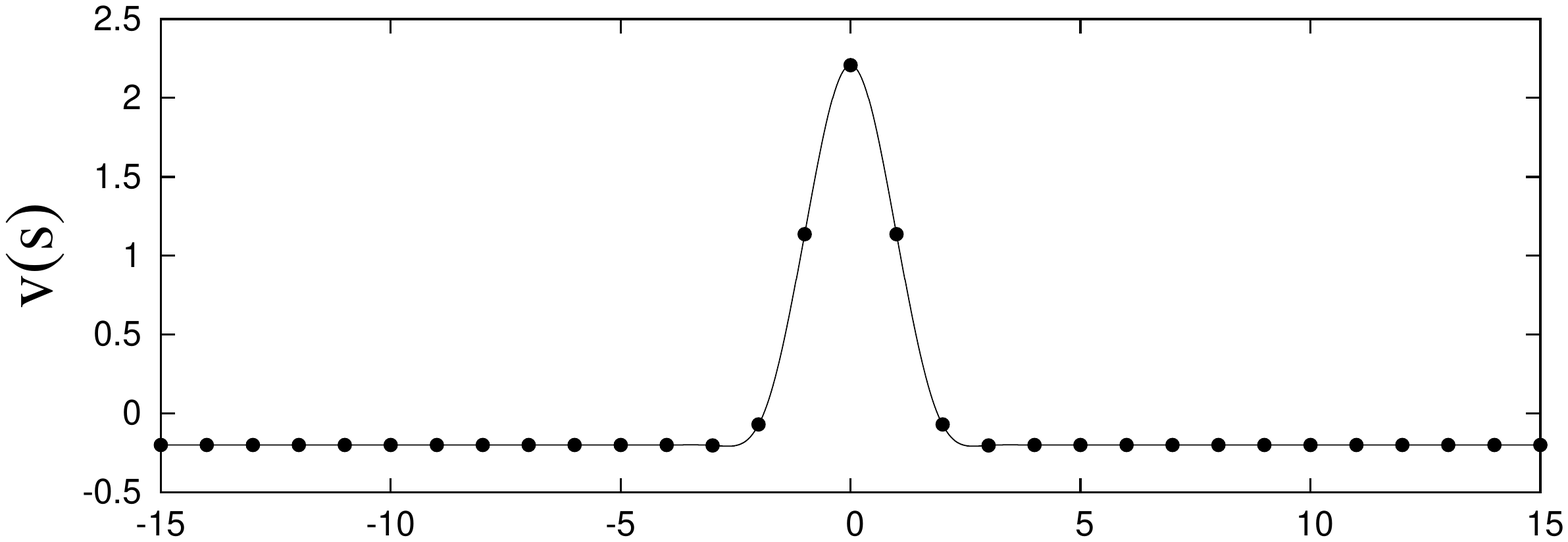}
    \includegraphics[draft=false,width=0.5\textwidth]{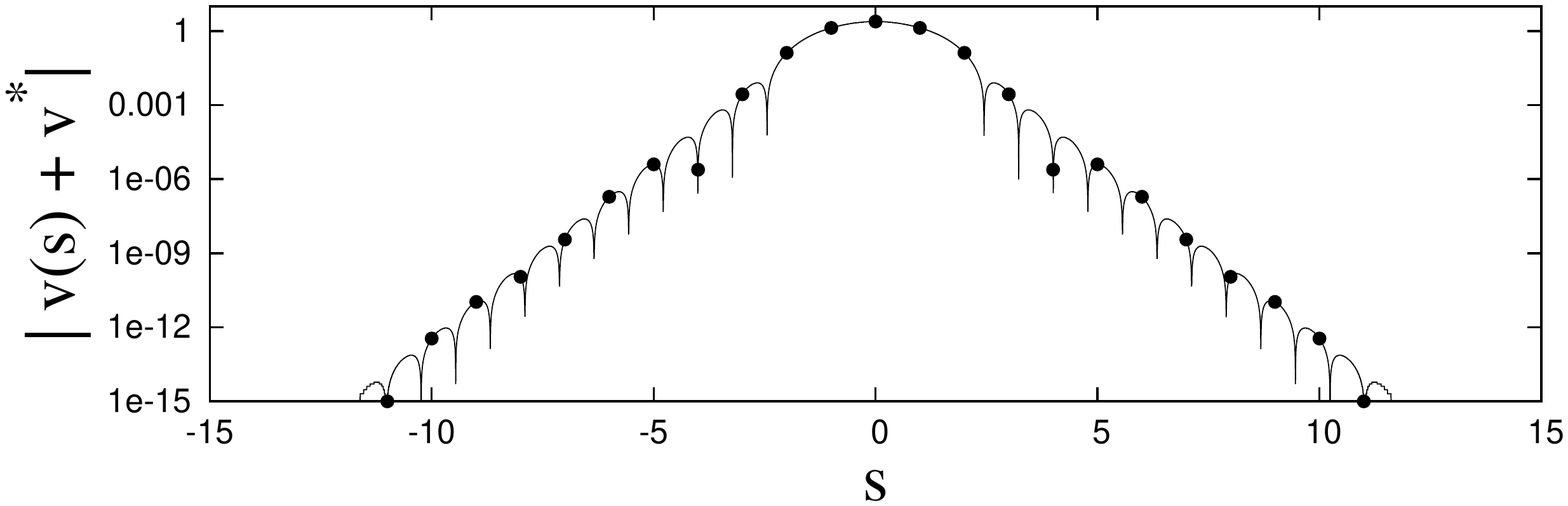}
  \end{center}
  \caption{Solitary wave with periodic and exponentially decaying
  tails. $v^*=-0.2$ and $\lambda=1.0$. The plot was generated with the help of
  the traveling wave scheme \eref{eq:lattice_twa_scheme} for the lattice
  equation; it has no QCA counterpart.}
  \label{fig:periodic_soliton}
\end{figure}

From the fixed point analysis of the advanced-delayed equation
\eref{eq:lattice_twa} a bifurcation occurs at point 2 in
Fig.~\ref{fig:ev}(a). So, if the velocity $\lambda$ reaches the critical point
$\lambda_{C}$, the fixed point changes its type from a center to a stable and
an unstable focus. This corresponds to a solitary wave with oscillatory
decaying tails. In Fig.~\ref{fig:periodic_soliton} we show an example of such
a wave. The offset is $v^*=-0.2$ and the wave velocity is $\lambda=1.0$. This
behavior does not occur in the quasi continuum.

\section{\label{sec:numerics}Numerical simulations of 
the initial value problem}

In this section we demonstrate, how the traveling waves described in the
previous sections appear in the course of the evolution of the lattice. We
will restrict ourselves to the simplest coupling term $q(v)=\cos v$, while the
background state will be general $v^* \neq 0$. Furthermore, we will study the
stability properties of the colliding waves.

\subsection{Evolution of an initial pulse}

First, we consider the evolution of an initial $\cos$-pulse with the coupling
$q(v)=\cos v$. The initial condition is
\begin{equation}
v_n(0) = 
\begin{cases}
v^* + \frac{A}{2} \left[1+\cos \left( \frac{n-n_0}{w} \pi \right) \right] &
|n-n_0|<w \\ v^* & \text{else,}
\end{cases}
\label{eq:lattice_ic}
\end{equation}
where $A$ is the amplitude, $n_0$ is the center and $w$ is the half width of
the pulse.

\begin{figure}
  \begin{center}
    \includegraphics[draft=false,width=0.5\textwidth]{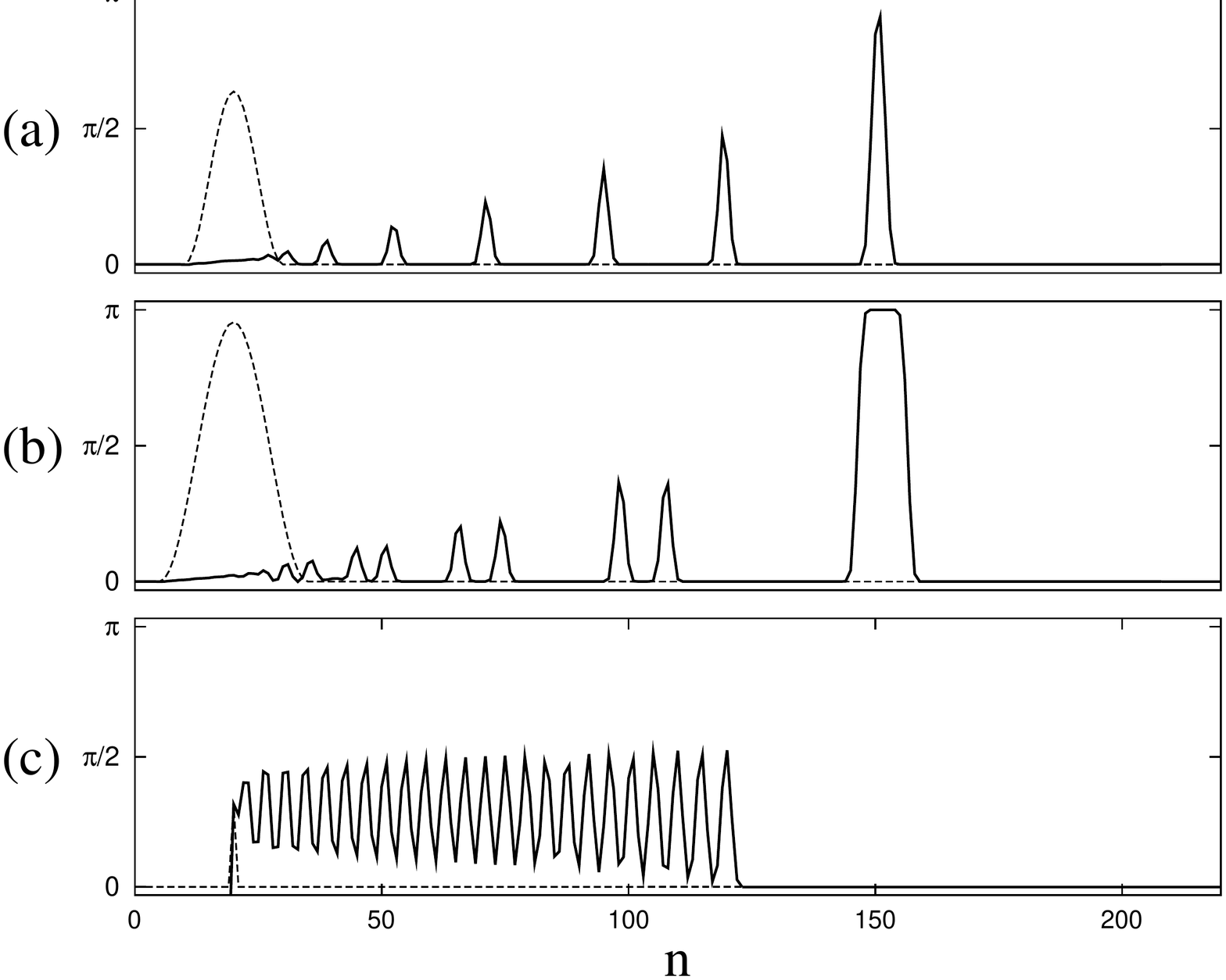}
  \end{center}
  \caption{Evolution of different initial pulses for the coupling $q(v)=\cos
  v$ and $v^*=0$. The initial conditions were set according to
  \eref{eq:lattice_ic}. (a) $w=10$ and $A=2$, (b) $w=15$ and $A=3$ and (c)
  $w=1$ and $A=1$. The dashed line shows the initial condition $v_n(t=0)$ and
  the solid line the lattice at the time $t=100$.}
  \label{fig:cos_pulse_pi0}
\end{figure}

\begin{figure}
  \begin{center}
    \includegraphics[draft=false,width=0.5\textwidth]{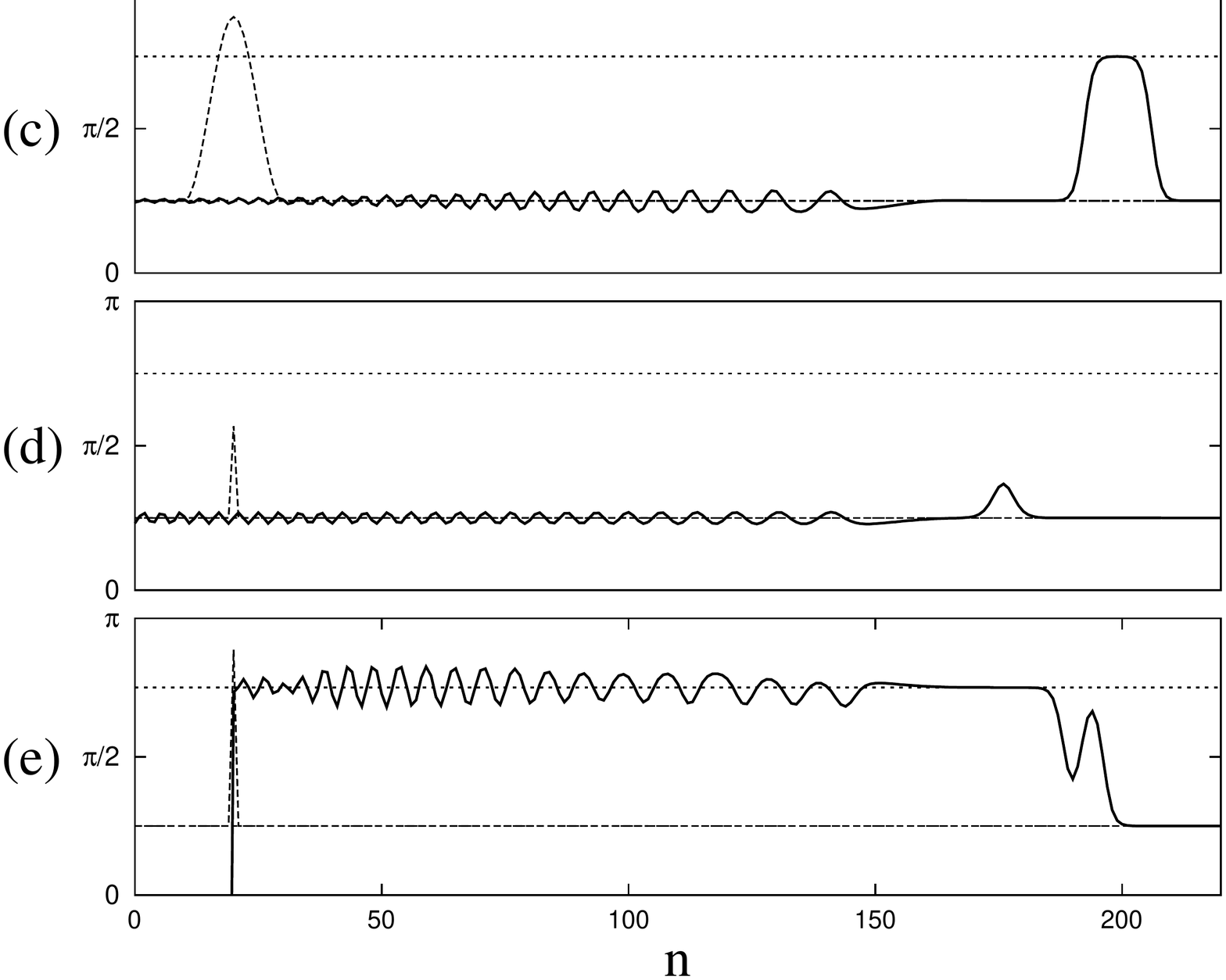}
  \end{center}
  \caption{Evolution of different initial pulses for the coupling $q(v)=\cos
  v$ and the background $v^*=\pi / 4$.  The initial conditions were set
  according to \eref{eq:lattice_ic}. (a) $w=10$ and $A=1$, (b) $w=15$ and
  $A=1.5$, (c) $w=10$ and $A=2$, (d) $w=1$ and $A=1$ and (e) $w=1$ and
  $A=2$. The dashed line shows the initial condition $v_n(t=0)$ and the solid
  line the lattice at the time $t=100$. Furthermore the position of the kink
  point at $3/4 \pi$ is shown.}
  \label{fig:cos_pulse_pi4}
\end{figure}

In Figs.~\ref{fig:cos_pulse_pi0},\ref{fig:cos_pulse_pi4} we compare the
evolution for different values of $v^*$.  In Fig.~\ref{fig:cos_pulse_pi0} we
set $v^*=0$ ($q'(v^*)=0$) and one can observe compactons and kovatons arising
from the initial pulse. In Fig.~\ref{fig:cos_pulse_pi0}(a) a wave train of
compactons emerges out of the initial pulse. The speed of the compactons
increases with increasing amplitude. In Fig.~\ref{fig:cos_pulse_pi0}(b) we
have increased the width and the amplitude of the pulse and one kovaton is
observed. In the bottom plot of Fig.~\ref{fig:cos_pulse_pi0} a narrow initial
pulse creates a wave source, emitting periodic waves.

In Fig.~\ref{fig:cos_pulse_pi4} we show results for the background
$v^*=\pi/4$. Here, $q'(v^*)\neq 0$ and the solitary waves arising from the
initial pulse possess exponential tails. Fig.~\ref{fig:cos_pulse_pi4}(a),(b)
are similar to Fig~\ref{fig:cos_pulse_pi0}(a),(b), where the initial pulse
decomposes into a train of solitons and kink. Note, that the number of emitted
solitary waves is smaller than for the case $v^*=0$. Furthermore, periodic
waves around $v^*$ can emerge, see Fig.~\ref{fig:cos_pulse_pi4}(c) and
(d). The plot in (d) is somehow similar to Fig.~\ref{fig:cos_pulse_pi0}(c),
with the difference, that appearing periodic waves are around
$v^*$. Fig.~\ref{fig:cos_pulse_pi4}(e) shows the evolution of a narrow initial
pulse with a relative large amplitude. It results in a kink with periodic
waves around the top of the kink. A detailed analysis of all possible
waveforms goes beyond the scope of this paper and will be reported elsewhere.

\subsection{Transition to chaos in a finite lattice}

\begin{figure}
  \begin{center}
    \includegraphics[draft=false,angle=270,width=0.95\textwidth]{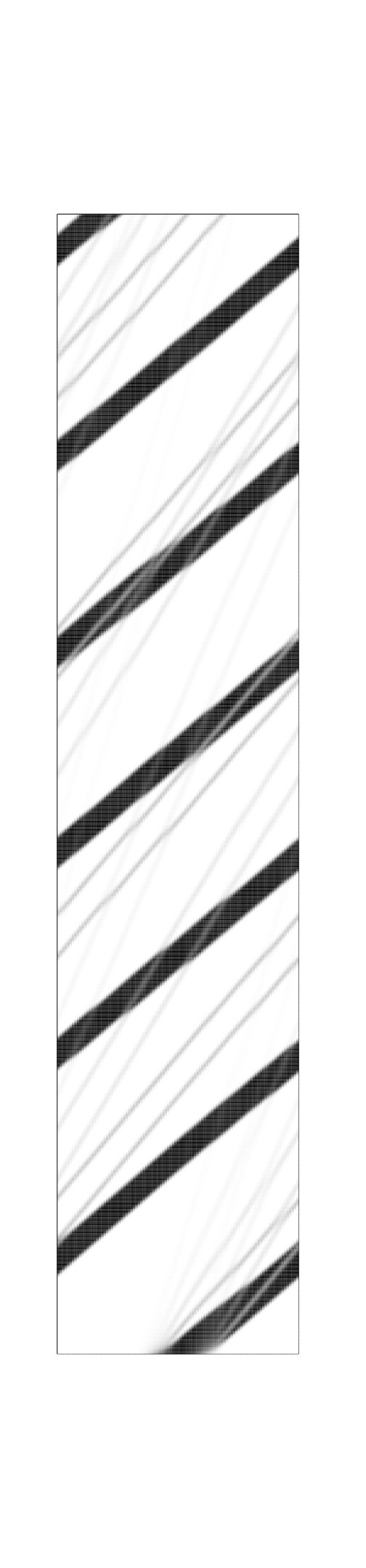}
    \includegraphics[draft=false,angle=270,width=0.95\textwidth]{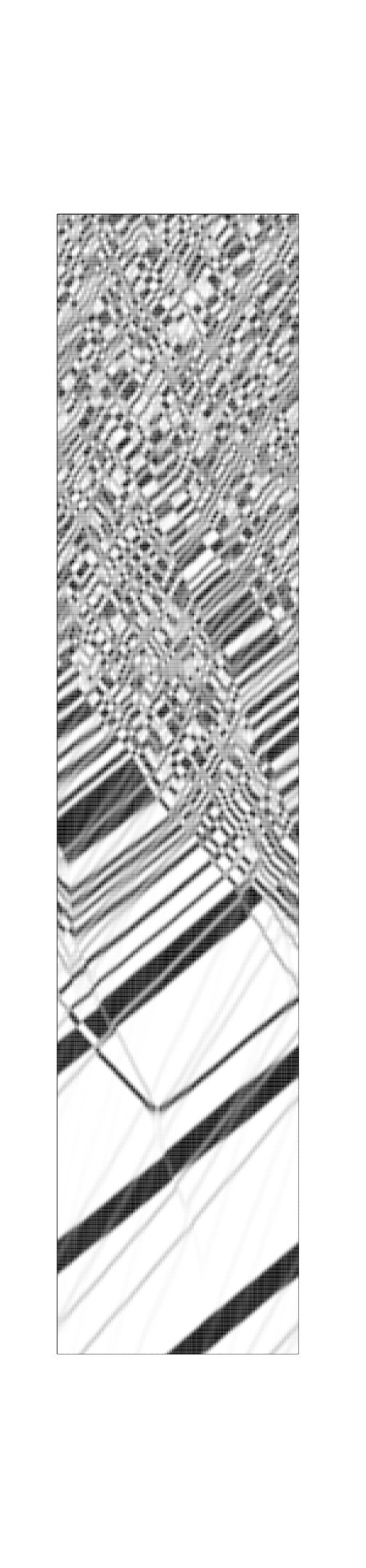}
  \end{center}
  \caption{Transition to chaos. The background is $v^*=0.1$ and the lattice
    contains of $N=100$ sites with periodic boundary conditions. The field is
    shown in a gray scale versus time (horizontal axis) and space (vertical
    axis). Upper plot: A kink and several solitons emerge out of an $\cos$
    pulse. The time interval is $0 \le t \le 400$. Lower plot: Emergence of
    chaos after a collision of two solitons which creates an
    soliton-antisoliton pair. The time interval is $2600 \le t \le 3000$.}
  \label{fig:transition_chaos}
\end{figure}

Wave trains shown in Figs.~\ref{fig:cos_pulse_pi0},\ref{fig:cos_pulse_pi4} are
obtain for an effectively infinite lattice (during the calculation times the
boundaries are not reached). In a finite lattice, collisions between waves
occur. We have used periodic boundary conditions, and observed that at large
times eventually a chaotic regime appears. In Fig.~\ref{fig:transition_chaos}
we show the evolution of an initial $\cos$ pulse with $v^*=0.1$. The upper
plot shows the initial decomposition of this pulse into one kink and several
solitary waves. These structures appear to survive collisions quite
unaffected. The lower plot shows, that after some transient time chaos
emerges. The chaotic state begins to develop, after a collision of two
solitons produces a large-amplitude soliton-antisoliton pair. Then an
avalanche of soliton-antisoliton collisions is triggered on, 
resulting in
a fast chaotization.

\begin{figure}
  \begin{center}
    \includegraphics[draft=false,width=0.5\textwidth]{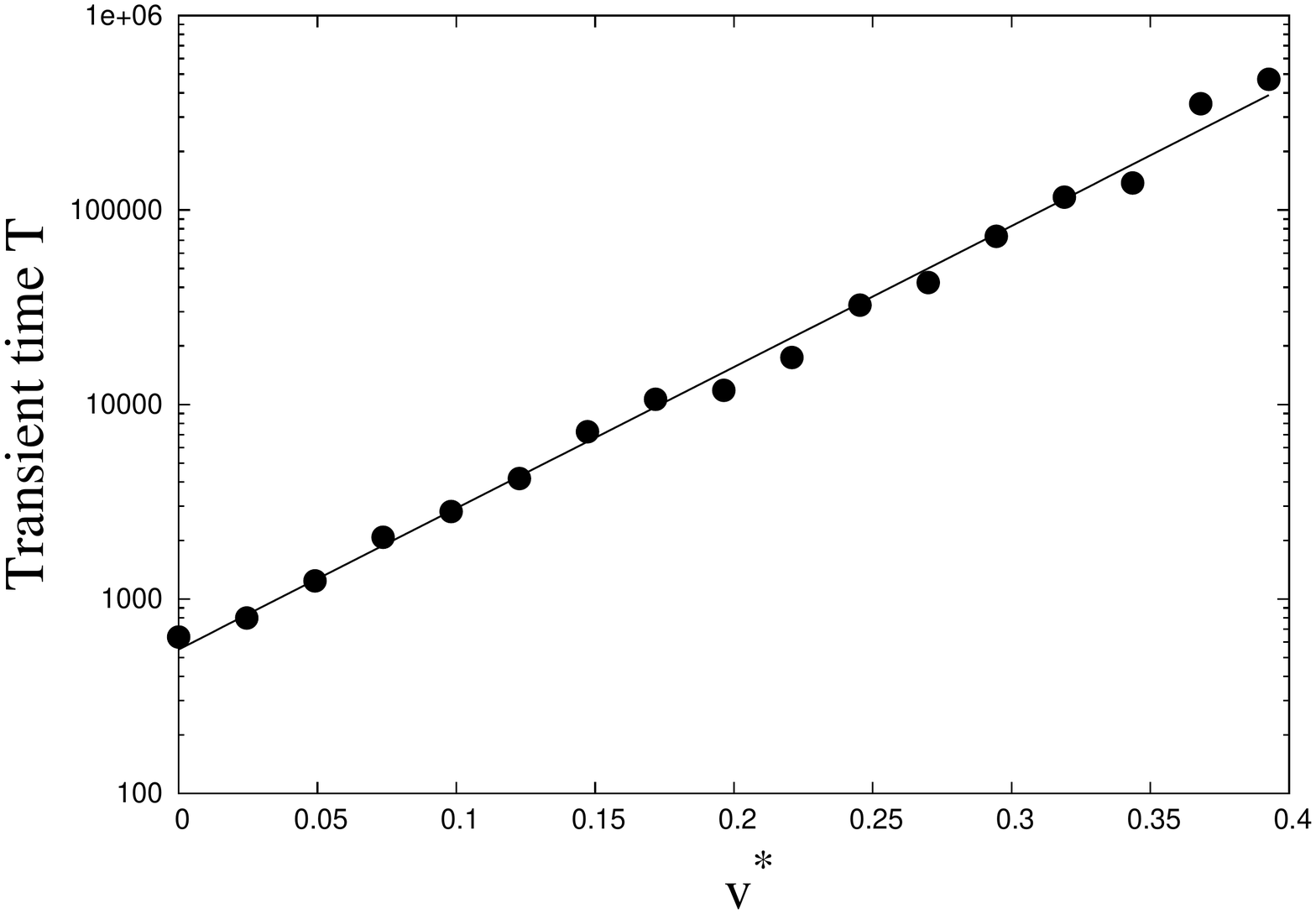}
  \end{center}
  \caption{Transient times to chaos for different backgrounds $v^*$. The
  length of the lattice is $N=32$ and the initial $\cos$-pulse is $v_n=\pi/4
  \big(\cos(\pi(n-N/2)/w)+1\big)+v^*$ for $|n-N/2|<w$ and $v=v^*$ else. In order
  to obtain an average of the transient times we also varied the width of
  the initial pulse from $5$ to $15$ and calculated the transient time as the
  average over the transient times for different initial pulses. The line
  is an exponential fit and the transient times scales with $T \sim \exp{ 16.7
  v^*}$.}
  \label{fig:stability}
\end{figure}

In Fig.~\ref{fig:stability} we show a remarkable dependence on the average
transient time, after which chaos establishes, on the parameter $v^*$. For
larger values of $v^*$ the transient time is exponentially large, what means
extreme stability of the solitary waves. Qualitatively, this stability can be
attributed to a smallness of effects of discreteness of the lattice for large
$v^*$. Here, the waves are relatively wide, thus they are well approximated in
the QCA, which is close to the integrable Korteveg-de Vries equation. For
small $v^*$ the waves are close to compactons that are short and for them the
discreteness that causes non-elasticity of collisions is
essential. Furthermore, the number of emitted waves decreases with increasing
$v^*$ and the velocity of the waves is bounded from below. These two effects
reduce the possibility that two waves meet each other, resulting in an
increased transient time.

\section{Conclusions}

We have demonstrated a variety of nontrivial wave structures in dispersively
coupled oscillator lattices. Remarkably, they appear in a very simple lattice
described by Eq.~\eref{eq:model}. In this study we have focused, contrary to
previous works~\cite{Rosenau-Pikovsky-05,Pikovsky-Rosenau-06}, on the features
that appear for a general, non-symmetric coupling function. While some
nontrivial solutions (compactons) survive in a general case, other (kovatons)
exist only in the symmetric situation. Instead, for a general case we have
reported a novel type of semi-compact waves. In our study of the waves on a
lattice we have described a novel transition from monotonic to oscillatory
tails of solitary waves that does not exist in the quasicontinuous
approximation.  Our comparison of general typical solutions of the lattice
model with special ones studied in
\cite{Rosenau-Pikovsky-05,Pikovsky-Rosenau-06} has shown that the waves with
exponential tails are much more ``resistant'' to chaotization compared to the
compactons.

Here, we would outline several possible
extensions of the analysis. In general, coupling between oscillators can
possess both dispersive and dissipative parts. The waves described in this
work will still be observed if the dissipation is sufficiently small, this is
confirmed by the perturbation analysis in \cite{Pikovsky-Rosenau-06}. Another
feature disturbing the waves is a non-homogeneity of the lattice, e.g. due to
non-uniformity of coupling. We expect that the waves will scatter on such
inhomogeneities, but this issue has not been studied yet. Finally, it is
intriguing, what kinds of waves can be observed in two- and three-dimensional
lattices; results in this direction will be published elsewhere.

We thank P. Rosenau for helpful discussions and DFG for support.

%\bibliographystyle{unsrt}
%\bibliography{nld-old,nld-current,%
%pap-ab,pap-ce,pap-fg,pap-hj,pap-kl,%
%pap-mn,pap-oq,pap-rs,pap-tz,%
%pik,books,n-stand}

\end{document}